\def\lesssim{{_ <\atop{^\sim}}}
\def\grtsim{{_ >\atop{^\sim}}}
\def\ap3m{AP$^3$M}
\def\LCDM{$\Lambda$CDM}
\def\LWDM{$\Lambda$WDM}
\def\hkpc{$h^{-1}{\ }{\rm kpc}$}
\def\hMpc{$h^{-1}{\ }{\rm Mpc}$}
\def\hMsun{$h^{-1}{\ }{\rm M_{\odot}}$}
\def\kms{${\rm{\ }km{\ }s^{-1}}$}
\def\nbody{$N$-body}
\def\c15{$c_{\rm 1/5}$}
\def\rvir{$r_{\rm vir}$}

\def\bx{{\bf x}}
\def\bu{{\bf u}}
\def\bw{{\bf w}}
\def\thr{\hat{\rho}}

\newcommand{\Eq}[1]{Eq.~(\ref{#1})}
\newcommand{\Eqs}[1]{Eqs.~(\ref{#1})}
\newcommand{\Fig}[1]{Fig.~\ref{#1}}
\newcommand{\mlapm}{\texttt{MLAPM}}
\newcommand{\mhf}{\texttt{MHF}}

\def\ea{et~al.~}                            
\def\o{\"o}                                 

\def\lesssim{\mathrel{\hbox{\rlap{\hbox{\lower4pt\hbox{$\sim$}}}\hbox{$<$}}}}
\def\gtrsim{\mathrel{\hbox{\rlap{\hbox{\lower4pt\hbox{$\sim$}}}\hbox{$>$}}}}

\newcommand{\AAA}[3]    {\mbox{A\&A~\textbf{#1},~#2~(#3)}}

\newcommand{\ApJ}[3]    {\mbox{ApJ~\textbf{#1},~#2~(#3)}}
\newcommand{\ApJS}[3]   {\mbox{ApJ~Suppl.~\textbf{#1},~#2~(#3)}}

\newcommand{\MNRAS}[3]  {\mbox{MNRAS~\textbf{#1},~#2~(#3)}}

\newcommand{\PhRevD}[3] {\mbox{Phys.~Rev.~\textbf{D#1},~#2~(#3)}}
\newcommand{\astroph}[1]{\mbox{\texttt{astro-ph/#1}}}

\newcounter{parentequation}
\newenvironment{subequations}{%
  \refstepcounter{equation}%
  \setcounter{parentequation}{\value{equation}}%
  \setcounter{equation}{0}%
  \ignorespaces
}{%
  \setcounter{equation}{\value{parentequation}}%
  \ignorespacesafterend
}

\documentstyle[epsfig]{mn}

\begin{document}

\title[Hydrodynamic Approch to Cosmic Structures]
      {Hydrodynamic Approach to the Evolution of Cosmic Structures II:
       Study of \nbody\ Simulations at $z=0$}
     
     \author[Knebe A. \ea] {Alexander Knebe$^{1}$, Alvaro
       Dom{\'\i}nguez$^2$ \&
       Rosa Dom{\'\i}nguez-Tenreiro$^3$\\
       {$^1$Astrophysikalisches Institut Potsdam, 
         An der Sternwarte 16, D-14482 Potsdam, Germany}\\
       {$^2$F{\'\i}sica Te{\'o}rica, Univ.\ de Sevilla, 
         Apdo.\ 1065, E-41080 Sevilla, Spain}\\
       {$^3$Departamento de F{\'\i}sica Te{\'o}rica, M{\'o}dulo C-XI,
         Univ.\ Aut{\'o}noma de Madrid, E-28049 Madrid, Spain}\\
     }

\date{Received ...; accepted ...}

\maketitle

\begin{abstract}
We present a series of cosmological \nbody\ simulations which make use of the hydrodynamic approach to the evolution of structures (Dom{\'\i}nguez 2000). This approach addresses explicitly the existence of a finite spatial resolution and the dynamical effect of subresolution degrees of freedom. We adapt this method to cosmological simulations of the standard \LCDM\ structure formation scenario and study the effects induced at redshift $z=0$ by this novel approach on the large--scale clustering patterns as well as (individual) dark matter halos.

Comparing these simulations to usual \nbody\ simulations, 
we find that (i) the new (hydrodynamic) model entails a proliferation of low--mass halos, and (ii) dark matter halos 
have a higher degree of rotational support. These results agree with the theoretical expectation about the qualitative behaviour of the "correction terms" introduced by the hydrodynamic approach: these terms act as a drain of inflow kinetic energy and a source of vorticity by the small--scale tidal torques and shear stresses.


\end{abstract}

\begin{keywords}
gravitation -- methods: numerical -- methods: $N$-body simulations -- galaxies: formation -- cosmology: theory
\end{keywords}

\section{Introduction}
Gravitational instability is commonly accepted as the basic mechanism for 
structure formation on large scales. Combined with the CDM model it leads 
to the picture of hierarchical clustering with wide support from deep galaxy 
and cluster observations. During the recent phase of cosmic evolution 
groups and clusters of galaxies condense from large scale density 
enhancements, and they grow by accretion and merger processes of the 
environmental cosmic matter.

But despite the fact that the currently favoured \LCDM\ model has
proven to be remarkably successful on large scales (cf. WMAP results,
Spergel et al. 2003, Spergel et al. 2006), recent high--resolution \nbody\ simulations still seem
to be in contradiction with observation on sub--galactic scales. 
There is, for instance, the problem with the steep central densities of galactic halos as the highest resolution simulations favor a cusp with
a logarithmic inner slope for the density profile of approximately
$-1.2$ (Diemand, Moore \& Stadel 2005; Fukushige, Kawai \& Makino 2004; Power~\ea 2003), whereas high resolution observations of low
surface brightness galaxies are best fit by halos with a core of
constant density (Simon~\ea 2005; de Block \& Bosma 2002; Swaters~\ea 2003). A further problem relates to the overabundance of small--sized (satellite)
halos; there are many more subhaloes predicted by cosmological simulations than actually observed in nearby galaxies (e.g., Moore et al.\ 1998, Klypin et al.\ 1999, Gottl\o ber et al.\ 
2003).  The lack of observational evidence for these
satellites has led to the suggestion that they are completely (or
almost completely) dark, with strongly suppressed star formation due
to the removal of gas from the small protogalaxies by the ionising
radiation from the first stars and quasars (Bullock et~al. 2000; Tully
et~al. 2002; Somerville 2002; Hoeft~\ea 2005). Other authors suggest that perhaps low mass
satellites never formed in the predicted numbers in the first place,
indicating problems with the \LCDM\ model in general, which is replaced
with Warm Dark Matter instead (Knebe~\ea 2002; Bode, Ostriker~\& Turok
2001; Col{\'\i}n~\ea 2000).  Suggested
solutions also include the introduction of self-interactions into
collisionless \nbody\ simulations (e.g.  Spergel~\& Steinhardt 2000;
Bento~\ea 2000), and
non-standard modifications to an otherwise unperturbed CDM power
spectrum (e.g.\ bumpy power spectra: Little, Knebe~\& Islam 2003;
tilted CDM: Bullock 2001c).
Recent results from (strong) lensing
statistics though suggest that the predicted excess of substructure is in
fact required to reconcile some observations with theory (Dahle~\ea
2003, Dalal~\& Kochanek 2002), but this conclusion has not been
universally accepted (Sand~\ea 2004; Schechter~\& Wambsganss 2002;
Evans~\& Witt 2003).

The discovery of the mismatch between observations and simulations is a
result of the increase in the resolution of \nbody\ simulations over
the last years.
This has emphasized the importance of the 
influence of subresolution scales on the simulated dynamics, at least
when it comes to halo properties.
The purpose of this work is to study
the hydrodynamic--like formulation of the formation of cosmologial
structure proposed recently by Dom{\'\i}nguez~(2000), dubbed SSE
(small--size expansion). The SSE addresses explicitly the existence of
a finite spatial resolution and the dynamical effect of subresolution
degrees of freedom. Although developed independently, the SSE approach
is close in spirit to the Large--Eddy Simulation of turbulent flow
(see, e.g., Pope 2000 and refs.\ therein). This is a method devised to
simulate only the relevant, large--scale degrees of freedom according
to the Navier--Stokes equations describing flow in the high--Reynolds
number (i.e., turbulent) regime: physically meaningful approximations
are made in order to model the coupling to the neglected,
small--scale degrees of freedom.

The SSE starts from the microscopic equations of motion for a set of
$N$ particles under their mutual gravity and provides a set of
hydrodynamic-like equations for the (coarse--grained) mass density and
velocity fields. These new equations now contain "correction terms"
which
describe the effects of the coarse--graining procedure and correct for
them, respectively.
It therefore only appears natural to implement these correction terms
into an (adaptive) mesh \nbody\ code where the density is treated in a
coarse--grained fashion, too: mesh--based Poisson solvers frequently used
for cosmological simulations smooth the particle distribution onto a
grid and hence deal with a coarse--grained density field when solving
for the potential via Poisson's equation. For this purpose we will adapt the open source
\nbody\ code \mlapm\ (Multi-Level Adaptive
Particle--Mesh)\footnote{Available at \texttt{http://www.aip.de/People/AKnebe/MLAPM}}.
The particles of the \nbody\ simulations presented in this study are effectively hydrodynamical
Lagrangian particles which move under the action not only of the
mesh--computed gravitational force, but also of the additional,
correction terms modeling sub--grid degrees of freedom in the context
of the SSE.
%

The rest of the work is structured as follows: in Sec.\ \ref{HAPPI} we
describe the theoretical background of the SSE and provide the link
to mesh--based \nbody\ codes. In Sec.\ \ref{Nbody} we present the
simulation of several models (standard \LCDM\ model, \LWDM\ model, and
two runs incorporating the SSE corrections). In Sec.\ \ref{Analysis}
we perform a comparative analysis of the four runs at redshift $z=0$ from two
complementary points of view: properties of the mass density
and velocity fields, and properties of halos. Finally, Sec.\ 
\ref{Conclusions} includes a discussion of the results and the
conclusions.

\section{The Hydrodynamic Approach} \label{HAPPI}

We deal with a system of nonrelativistic, identical point particles
which (i) are supposed to interact with each other via gravity only,
(ii) look homogeneously distributed on sufficiently large scales, so
that the evolution at such scales corresponds to an expanding
Friedmann-Lema\^\i tre cosmological background, and (iii) deviations
to homogeneity are relevant only on scales small enough that a
Newtonian approximation is valid to follow their evolution. Let $a(t)$
then denote the expansion factor of the Friedmann-Lema\^\i tre
cosmological background, $H(t) = \dot{a}/a$ the associated Hubble
function, and $\varrho_b (t)$ the homogeneous (background) density on
large scales.  ${\bf x}_\alpha$ is the comoving spatial coordinate of
the $\alpha$-th particle, ${\bf u}_\alpha=a \dot{\bf x}_\alpha$ its
peculiar velocity, and $m$ its mass. In terms of these variables the
evolution is described by the following set of equations (Peebles
1980) ($\nabla_\alpha$ denotes a partial derivative with respect to
${\bf x}_\alpha$):
  \begin{subequations} \label{newton}
    \begin{eqnarray}
     \dot{\bf x}_\alpha & = & {1 \over a} {\bf u}_\alpha , \\
     \dot{\bf u}_\alpha & = & {\bf w}_\alpha - H {\bf u}_\alpha , \\
     \nabla_\alpha \cdot {\bf w}_\alpha & = & - 4 \pi G a \left[ {m \over a^3} 
      \sum_{\beta \neq \alpha} \delta^{(3)}({\bf x}_\alpha - {\bf x}_\beta) - \varrho_b \right] , \\
     \nabla_\alpha \times {\bf w}_\alpha & = & {\bf 0} ,
    \end{eqnarray}
  \end{subequations}
\noindent
where ${\bf w}_\alpha$ is the peculiar gravitational acceleration acting on
the $\alpha$-th particle. Finally, \Eqs{newton} must be subjected
to periodic boundary conditions in order to yield a Newtonian
description consistent with the Friedmann-Lema\^\i tre solution on
large scales (Buchert~\& Ehlers 1997).

If we now assume that the actual measure of the density field in
an \nbody\ code depends on a smoothing window $W(z)$, the microscopic
field $\varrho_{mic}$ relates to the measured (coarse--grained) field $\varrho$ in the following way:
  \begin{subequations}  \label{density}
    \begin{eqnarray}
     \varrho_{mic} ({\bf x}, t) & = & {m \over a(t)^3} \sum_\alpha 
    \; \delta^{(3)} ({\bf x}-{\bf x}_\alpha(t)) , \\
     \varrho ({\bf x}, t; L) & = & \int {d{\bf y} \over L^3} \; 
    W \left( {|{\bf x} - {\bf y}| \over L} \right) 
    \varrho_{mic} ({\bf y}, t) .
   \end{eqnarray}
  \end{subequations}
\noindent
The physical interpretation of the field $\varrho ({\bf x}; L)$
follows straightforwardly from the properties of the smoothing window:
it is proportional to the number of particles contained within the
coarsening cell of size $\approx L$ centered at ${\bf x}$. A
microscopic peculiar--momentum density field and the corresponding
coarse--grained field can be defined in the same way:
  \begin{subequations} \label{momentum}
   \begin{eqnarray}
    {\bf j}_{mic} ({\bf x}, t) & = & {m \over a(t)^3} 
    \sum_\alpha {\bf u}_\alpha(t) \; \delta^{(3)} ({\bf x}-{\bf x}_\alpha(t)) , \\
    {\bf j} ({\bf x}, t; L) & = & \int {d{\bf y} \over L^3} \; 
    W \left( {|{\bf x} - {\bf y}| \over L} \right) 
    {\bf j}_{mic} ({\bf y}, t) .
   \end{eqnarray}
  \end{subequations}
\noindent
One can introduce peculiar velocity fields ${\bf u}_{mic}$ and ${\bf
  u}$ by definition as ${\bf j} =: \varrho \, {\bf u}$ and similarly
for ${\bf u}_{mic}$. The physical meaning of ${\bf u} ({\bf x}; L)$ is
also simple: it is the center-of-mass peculiar velocity of the
subsystem defined by the particles inside the coarsening cell of size
$\approx L$ centered at ${\bf x}$. Notice that ${\bf u}$ is {\em not}
obtained by coarse-graining ${\bf u}_{mic}$: from a dynamical point of
view, it is more natural to coarse--grain the momentum rather than the
velocity, since the former is an additive quantity for a system of
particles. Finally, one can define peculiar gravitational acceleration
fields ${\bf w}_{mic}$ and ${\bf w}$ through coarse--graining of the
force:
\begin{subequations} \label{acceleration}
  \begin{eqnarray}
    \varrho_{mic} {\bf w}_{mic} ({\bf x}, t) & = & {m \over a(t)^3} 
    \sum_\alpha {\bf w}_\alpha(t) \; \delta^{(3)} ({\bf x}-{\bf x}_\alpha(t)) ,\\
    \varrho \, {\bf w} ({\bf x}, t; L) & = & \int {d{\bf y} \over L^3} \; 
    W \left( {|{\bf x} - {\bf y}| \over L} \right) 
    \varrho_{mic} {\bf w}_{mic} ({\bf y}, t) .
  \end{eqnarray}
\end{subequations}
\noindent
The field ${\bf w} ({\bf x})$ has the physical meaning of the
center-of-mass peculiar gravitational acceleration of the subsystem
defined by the coarsening cell at ${\bf x}$.

From these definitions and \Eqs{newton}, it is straightforward
to derive the evolution equations obeyed by the coarse--grained fields
$\varrho$ and ${\bf u}$ (from now on, $\partial/\partial t$ is taken
at constant ${\bf x}$ and L, and $\nabla$ means partial derivative with
respect to ${\bf x}$):
\begin{subequations}  \label{hydro}
  \begin{equation}
    \frac{\partial \varrho}{\partial t} + 3 H \varrho = - {1 \over a} 
    \nabla \cdot (\varrho \, {\bf u}) ,
  \end{equation}
  \begin{equation}
    \frac{\partial (\varrho \, {\bf u})}{\partial t} + 4 H \varrho \, 
    {\bf u} = \varrho \, {\bf w} - 
    {1 \over a} \nabla \cdot (\varrho \, {\bf u} \, {\bf u} + \Pi) ,
  \end{equation}
\end{subequations}
\noindent
where a new second-rank tensor field has been defined (dyadic notation):
\begin{eqnarray}  \label{veldis}
  \Pi ({\bf x}, t; L) & = & \int {d{\bf y} \over L^3} \; 
  W \left( {|{\bf x} - {\bf y}| \over L} \right) \varrho_{mic}({\bf y}, t) \\
  & & \!\!\!\! [{\bf u}_{mic}({\bf y}, t) - {\bf u}({\bf x}, t; L)]
  [{\bf u}_{mic}({\bf y}, t) - {\bf u} ({\bf x}, t; L)] . \nonumber
\end{eqnarray}

\noindent
The field $\Pi ({\bf x})$ is due to the velocity dispersion, i.e., to
the fact that the particles in the coarsening cell have in general a
velocity different from that of the center of mass. 
%
The physical meaning of \Eqs{hydro} is simple: they are just balance
equations, stating mass conservation and momentum conservation,
respectively. The term $\varrho \, {\bf w}$ codifies the gravitational
interaction between the coarsening cells and does not satify, in
general, Poisson's equation or the curl--free condition. The term
$\nabla \cdot \Pi$ represents a {\em kinetic} pressure force due to
the exchange of particles between neighboring coarsening cells (just
like for an ideal gas) and it has the same physical origin as the
convective term $\nabla \cdot (\varrho \, {\bf u} \, {\bf u})$, i.e.,
a nonlinear mode-mode coupling of the velocity field. The difference
is that the convective term couples only modes on scales $> L$, while
the velocity dispersion term codifies the dynamical effect of the
coupling of the modes on scales $> L$ with the modes on scales $< L$.
\Eqs{hydro} are {\em exact}: as one changes the smoothing length, the
fields $\varrho$, ${\bf u}$, ${\bf w}$, $\Pi$ change in such a way
that the form of the equations remains the same (for example, upon
increase of the smoothing length, part of the dynamical effect
described by $\nabla \cdot (\varrho \, {\bf u} \, {\bf u})$ is shifted
towards $\nabla \cdot \Pi$). This property is reflected in that the
equations are not an autonomous system for $\varrho$ and ${\bf u}$. In
fact, they are just the first ones of an infinite hierarchy, as can be
checked by computing the evolution equations for the fields ${\bf w}$
and $\Pi$.  To obtain a useful set of equations, it is necesary to
truncate this hierarchy by looking for a functional dependence of
${\bf w}$ and $\Pi$ on $\varrho$ and ${\bf u}$.

The peculiarities of the problem at hand (collisionless matter in the
non--stationary state of structure formation) prevent the usual
truncation of the hierarchy leading to the Euler or Navier--Stokes
equations, respectively (see, e.g., Chapman \& Cowling 1991).
The \textit{small--size expansion} (SSE) is a specific truncation for
this problem (Dom{\'\i}nguez 2000, 2002; Buchert \& Dom{\'\i}nguez
2005), that starts from the physical assumption that the coupling to
the small--scales is weak (this can be argued on the basis that, in a
hierarchical scenario, the smaller scales ``virialize'' earlier and
thus ``decouple'' from the evolution of the larger scales). Then the
fields $\Pi$ and $\bw$ are derived as a formal expansion in $L$:
Keeping terms up to order $L^2$, \Eqs{hydro} become ($\partial_i =
\partial/\partial x_i$; summation over the repeated index $i$ is
understood)
\begin{subequations}
  \label{hydrosmallk}
  \begin{equation}
    \label{sse_mass}
    \frac{\partial \varrho}{\partial t} + 3 H \varrho = - {1 \over a} 
    \nabla \cdot (\varrho \, {\bf u}) ,
  \end{equation}
  \begin{equation}
    \label{sse_momentum}
    \frac{\partial (\varrho \, {\bf u})}{\partial t} + 4 H \varrho \, 
    {\bf u} = \varrho \, {\bf w}^{\rm mf} - 
    {1 \over a} \nabla \cdot (\varrho \, {\bf u} \, {\bf u})
    + \varrho \, {\bf C} ,
  \end{equation}
  \begin{equation}
    \label{sse_poisson}
    \nabla \cdot {\bf w}^{\rm mf} = - 4 \pi G a \, ( \varrho - \varrho_b ) ,
  \end{equation}
  \begin{equation}
    \label{sse_nocurl}
    \nabla \times {\bf w}^{\rm mf} = {\bf 0} ,
  \end{equation}
\end{subequations}

\noindent
with the additional acceleration
\begin{equation} \label{correction}
 {\bf C} = \frac{B L^2}{\varrho} 
                  \left[
                      (\nabla \varrho \cdot \nabla) {\bf w}^{\rm mf} -
                      {1 \over a} \nabla \cdot [\varrho (\partial_i {\bf u}) 
      (\partial_i {\bf u})]  
                  \right] .
\end{equation}
The constant $B$ is determined by the smoothing window $W(z)$,

\begin{equation} \label{B}
 B = {1 \over 3} \int d{\bf z} \; z^ 2 \, W(z) = {4 \pi \over 3} 
  \int_0^{+\infty} d z \; z^4 \, W(z) .
\end{equation}

\noindent 
To order $L^0$, \Eqs{hydrosmallk} reduce to the "dust (pressureless)
approximation" for cosmological structure formation (Sahni \& Coles
1995): ${\bf w}^{\rm mf}$ represents the {\it mean--field} gravity
created by the monopole moment of the matter distribution in the
coarsening cells, i.e., the total mass, and the coupling to the
small--scales is neglected altogether. To order $L^2$ there are two
kinds of corrections: the term proportional to the tidal tensor
$\nabla {\bf w}^{\rm mf}$ (stemming from a term $\bw-\bw^{\rm mf}$ in
\Eqs{hydro}) models the gravitational force of the higher--order
multipole moments, i.e., the coupling to the subresolution {\em
  configurational} degrees of freedom; the term proportional to
$(\partial_i {\bf u}) (\partial_i {\bf u})$ (stemming from $\Pi$ in
\Eqs{hydro}) models the effect of velocity dispersion, that is, the
coupling to the subresolution {\em kinetic} degrees of freedom.

The expected dynamical effect of these new terms 
with respect to the ``dust evolution'' has been studied theoretically
(Dom{\'\i}nguez 2000, 2002; Buchert \& Dom{\'\i}nguez 2005).
There is evidence that, assuming a locally
plane--parallel collapse, these terms mimick the ``adhesion model''
(Gurbatov, Saichev~\& Shandarin 1989, Sahni \& Coles 1995), in which
recently collapsed regions stabilize --- or more generally speaking, the term
due to velocity dispersion tends to reduce the inflow velocity in
collapsing regions and favors the formation of gravitationally
  bound systems. 
It has also
been shown that the correction terms act as a source of vorticity
via small--scale tidal forces and shear stresses.
The ''dust model'' lacks a source of vorticity $\bomega = \nabla
\times \bu$, and the initially present one is damped by the cosmological
expansion in the linear regime. Thus, in that respect the corrections to the ''dust
model'' can be particularly important. Taking the curl
of \Eq{sse_momentum} we obtain 
\begin{equation} 
  \label{vorticity} 
  \frac{\partial \bomega}{\partial t} 
  = \mbox{} - H \bomega + \frac{1}{a} \nabla \times (\bu \times \bomega)
  + \nabla \times {\bf C} \ ,
\end{equation}
where the term $\nabla\times{\bf C}$ is a source of vorticity, i.e., it does
not vanish in general even if $\bomega={\bf 0}$, as has been confirmed
perturbatively by Dom{\'\i}nguez (2002).
We comment further on the relationship between vorticity and angular momentum in
App.~\ref{sec:conservation}, where also some results concerning the
conservation of energy, momentum, and angular momentum are derived.
%
%
%

The dynamical evolution predicted by \Eqs{hydrosmallk} can be
implemented without much difficulties in a particle--mesh (PM) code of
\nbody\ simulation.
Mass conservation, \Eq{sse_mass}, is automatically satisfied by the
code. The acceleration $\bw^{\rm mf}$ given by \Eqs{sse_poisson} and
(\ref{sse_nocurl}) agrees with the value returned by the Poisson
solver on a grid, and the grid constant sets naturally the resolution
$L$.
In principle, one only needs to take care of \Eq{sse_momentum}, which
can be re-written in Lagrangian coordinates as 

\begin{equation} \label{dotu}
  \dot{\bf u} = {\bf w}^{\rm mf} - H {\bf u} + {\bf C} .
\end{equation}
The computation of ${\bf C}$, given by \Eq{correction}, on the grid of
the Poisson solver is a highly non-trivial but manageable task. Thus,
\Eq{dotu} --- together with $\dot{\bf x} = {\bf u}/a$, see Eq.~(\ref{newton}a) --- determines
the motion of Lagrangian fluid elements, which are sampled by the
particles of the simulation.  If we set $B=0$ ($\Rightarrow {\bf
  C}={\bf 0}$) in \Eq{dotu} we recover the equations of motions that
are being integrated in a standard PM code for the update of particle
velocities and positions during the course of the simulation.

\section{The $N$-body Simulations} \label{Nbody}
\subsection{The Setup}\label{setup}
The \nbody\ simulations presented in this study were carried out
using a version
 of the open source adaptive mesh
refinement code \mlapm\ (Knebe, Green~\& Binney 2001).
%
This code reaches high force resolution by refining high-density regions with an
automated refinement algorithm.  These adaptive meshes are recursive:
refined regions can also be refined, each subsequent refinement having
cells that are half the size of the cells in the previous level.  This
creates a hierarchy of refinement meshes of different resolutions
covering regions of interest.  The refinement is done cell by cell
(individual cells can be refined or de-refined) and meshes are not
constrained to have a rectangular (or any other) shape. The criterion
for (de-)refining a cell is simply the number of particles within that
cell and a detailed study of the appropriate choice for this number
can be found elsewhere (Knebe et al.\ 2001). The code also uses
multiple time steps on different refinement levels where the time step
for each level is two times smaller than the step on the previous
level. The latest version of \mlapm\ also includes an adaptive time
stepping that adjusts the actual time step after every major step to
restrict particle movement across a cell to a particular fraction of
the cell spacing, hence, fine tuning accuracy and computational time.

As outlined above, the only necessary modification required to model
the "\textbf{H}ydrodynamic \textbf{APP}rox\textbf{I}mation" (or
"HAPPI" here afterwards) is to account for the correction term ${\bf
  C}$ in \Eq{dotu} when updating the particle velocities\footnote{The
  modifications are part of \mlapm\ v1.4 (and all later versions) and can be switched on using
  \texttt{-DHAPPI} upon compile time.}. \mlapm\ has therefore been
modified to not only calculate the density field on its hierarchy of
nested refinement grids but also the velocity field. 
The $\nabla$-operator and spatial derivatives, respectively, have been
approximated by finite-differences using the two nearest neighbors (in
each dimension) to the cell for which the correction term is being
calculated. Cells close to a refinement boundary for which not enough
surrounding nodes are present, obtain their correction values
interpolated downwards from the next coarser grid level.
The assignment of mass and momentum on the grid is done with a
triangular--shaped--cloud window (Hockney~\& Eastwood 1988),

\begin{equation} \label{TSC}
 W(z) = \left\{
 \begin{array}{ll}
   \frac{3}{4} - z^2                              & \mbox{\rm for } |z|\leq 0.5 , \\
   \frac{1}{2} \left( \frac{3}{2} - |z| \right)^2 & \mbox{\rm for } 0.5<|z|\leq 1.5 , \\
   0                                              & \mbox{\rm otherwise},  \\
 \end{array}
 \right.
\end{equation}
for which $B=1/4$ according to \Eq{B}. We remark that, due to the
dynamical (de)refinement procedure of the \mlapm\ code, the resolution
is space--dependent in a discrete manner, while \Eqs{hydrosmallk} were
derived under the assumption of a spatially homogeneous length $L$. The SSE can
be generalized to the case of a smoothly varying $L(\bx)$
(Dom{\'\i}nguez, unpublished; Dom{\'\i}nguez 2002 contains the
generalization to a time--dependent $L$), but we have neglected this
additional complication because 
the fraction of particles which are in regions where $L$ jumps is less
than $1\%$ during the run.
Finally, we also note that this numerical method of integrating the
hydrodynamic equations is different from, albeit similar to, the
Smoothed Particle Hydrodynamics method (Gingold \& Monaghan 1977; Lucy
1977) frequently used in cosmological simulations involving baryonic
matter.

We ran four CDM simulations with cosmological parameters in
agreement with the so-called concordance model, i.e. $\Omega_0 = 1/3$,
$\lambda_0 = 2/3$, $\sigma_8=0.88$, $h=2/3$:

\begin{itemize}
 \item one standard \LCDM\ model,
 \item one \LWDM\ model ($m_{\rm WDM} = 0.5$keV),
 \item two \LCDM\ models with $B$=1/4 and $B$=1, respectively.
\end{itemize}
Even though $B$ is actually determined by the smoothing window, we
also considered a four times larger value as if it were a free parameter
of the model. This model is to be understood
as an "academic toy model" where we hope to gain better insight into the
effects of the correction term on cosmological structure formation.

All simulations consist of $N=128^3$ particles in a box of side length
25\hMpc\ (the mass of a simulation particle is $m_{\rm part} \approx 7
\times 10^8$\hMsun), and they were started at redshift $z=35$. The two
\LCDM\ models with $B\neq$0 are also dubbed \LCDM happi1 ($B$=1/4) and
\LCDM happi2 ($B$=1). We chose to also run a \LWDM\ model to allow for
a more complete comparision of the new HAPPI models to other,
alternative cosmologies. A more elaborate study of the \LWDM\ model
and Warm Dark Matter can be found in Knebe~\ea (2002).

The force resolution in \mlapm\ is determined by the finest refinement
level reached throughout the run. While all four models applied
exactly the same refinement criterion (six particles per cell),
the \LCDM happi2 run only invoked five refinement levels whereas all
other runs used seven levels at redshift $z=0$. In terms of force
resolution this translates to 10\hkpc\ spatial resolution for
\LCDM happi2 (corresponding to an estimated maximum density $\sim 5 \times 10^4 \varrho_b$) and 2.5\hkpc\ for all the other models (maximum density $\sim 3\times 10^6 \varrho_b$).
This difference can be ascribed to the smoothening effect of the correction
terms in the dynamical equations that are more effective for higher
values of $B$.

   \begin{figure}
      \centerline{\psfig{file=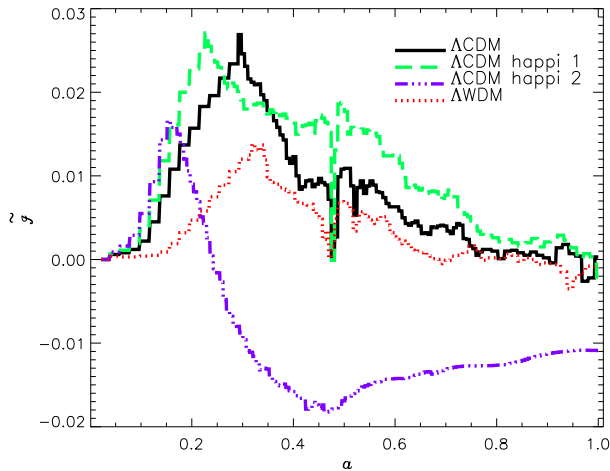,width=\hsize}}
      \caption{Variation with the expansion factor $a$ of the dimensionless quantity
        $\tilde{\cal I}$ defined by \Eq{eq:LI}.}
      \label{accuracy}
   \end{figure}

\subsection{Accuracy of the code} \label{LayzerIrvine}
A useful check of the accuracy of a simulation is
provided by the global invariants of the dynamical system, which we
derive and discuss in App.~\ref{sec:conservation}.

The dynamical evolution conserves the quantity $a \sum_\alpha \bu_\alpha$
(if the force and density interpolation schemes are identical). It was found
that departures from the initially vanishing value satisfy the bound

\begin{equation}
  \frac{1}{N}\left|\sum_{\alpha=1}^N \bu_\alpha\right| \lesssim 
  3 \times 10^{-3} \, u_{\rm mean} , \qquad 
  u_{\rm mean} = \frac{1}{N} \sum_{\alpha=1}^N |\bu_\alpha| \ ,
\end{equation}
where at redshift $z=0$ the average particle velocity was $u_{\rm mean}
\approx 300$\kms\ in all four models.

In standard $N$--body codes it is common practice
to check constancy of the invariant ${\cal I}$ (see \Eq{eq:defI})
following from the Layzer--Irvine equation (e.g., Knebe et al. 2001). This is not the case for the two
HAPPI models, since the correction term ${\bf C}$ is a source or drain
of energy as discussed in the App.~\ref{sec:conservation}.
We though chose to plot in \Fig{accuracy} the dimensionless quantity
\begin{equation}
  \label{eq:LI}
  \tilde{\cal I}(t) = 
  \frac{{\cal I}(t) - {\cal I}(t_1)}{a U^{\rm mf} (t)} ,
\end{equation}
as a function of cosmic expansion factor $a$, where $U^{\rm mf}$ is the mean--field potential energy defined in
\Eqs{eq:KU}. For the \LCDM\ and \LWDM\ models, the Layzer--Irvine
equation holds and predicts $\tilde{\cal I}=0$. Departures from this
result are ascribed to both integration/truncation errors introduced by the code and the fact that the particle shape is neither constant in time nor space due to the adaptive nature of \mlapm\ (cf. Knebe et al.\ 2001). \LCDM
happi1 performs rather similar to the standard CDM model, indicating
that the effect of the correction term ${\bf C}$ in the evolution of
${\cal I}$ is small compared to the numerical errors. \LCDM happi2,
on the other hand, departs noticeably from the other models and
$\tilde{\cal I}$ changes sign at around a redshift of $z\approx 3$,
which lets us expect to find larger differences between \LCDM happi2 and \LCDM.

\subsection{The Importance of the HAPPI correction} \label{HAPPIimportance}
\begin{figure*}
     \centerline{\psfig{file=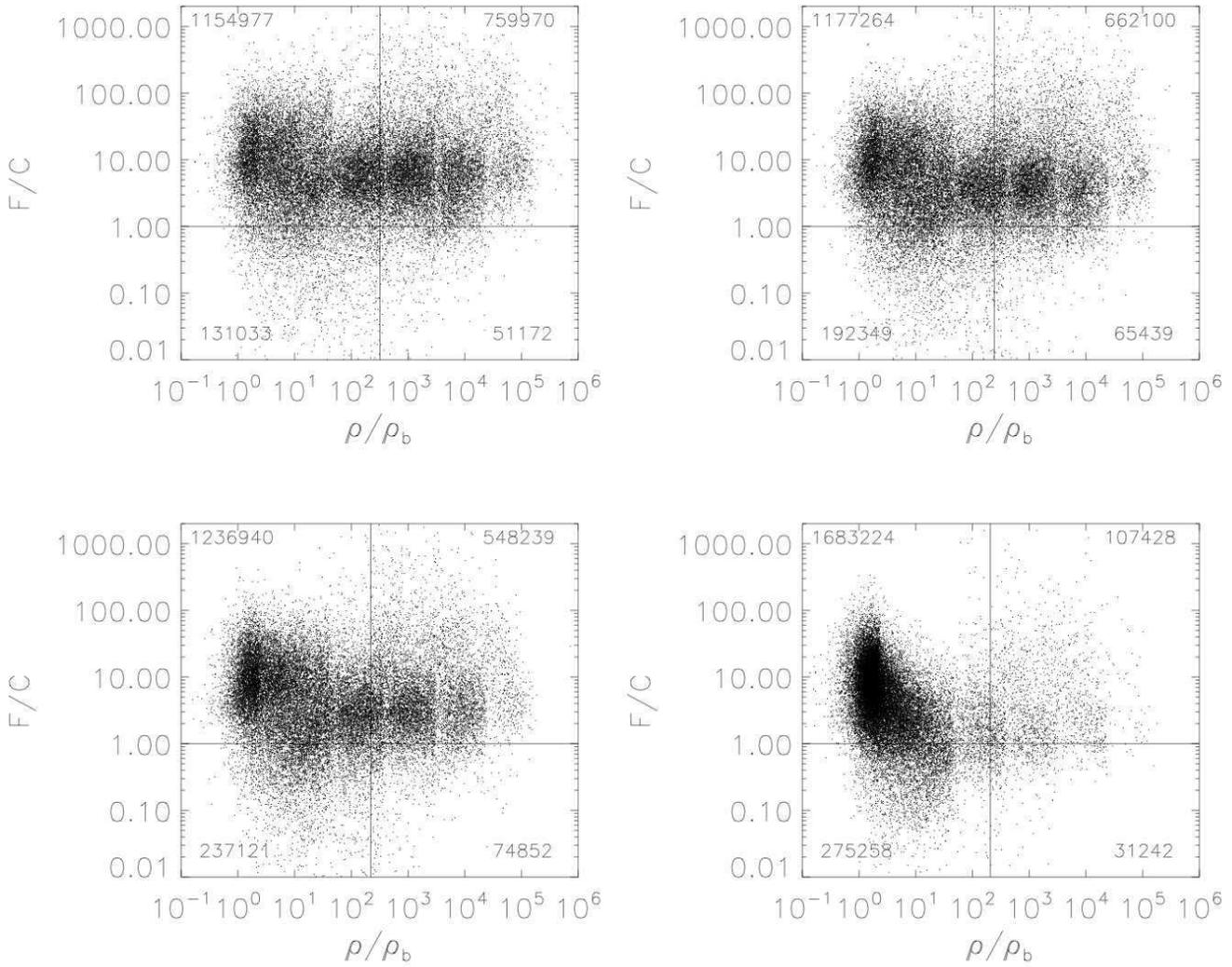,width=\hsize}}
      \caption{The correction term in comparison to the mean--field
        force acting on a single particle as a function of density.
        The four panels are for all particles in the simulation at
        redshifts $z=0$ (upper left), $z=0.5$ (upper right), $z=1.0$
        (lower left), and $z=4.0$ (lower right). 
        The vertical line indicates the virial overdensity at the
        respective redshift. The number in each quadrant lists the
        total number of particles in this regime of the plot.}
      \label{HAPPIdens}
   \end{figure*}

In order to test the importance of the correction term \Eq{correction}
we calculated the ratio of the mean--field acceleration (i.e.\
$F=|{\bf w}^{\rm mf}|$) and the additional acceleration ($C=|{\bf
  C}|$) for each individual particle as a function of the local
density at various redshifts. The result for the \LCDM happi2 model,
for which the effect of ${\bf C}$ is the largest, can be viewed in
\Fig{HAPPIdens}. This figure indicates that the effect of the newly
added terms is generally rather small especially at late times. At
redshift $z=0$ the fraction of all particles with $F/C<1$ ("HAPPI
particles") is a mere 9\% while it increases to 15\% at $z=4$.
In order to examine a possible trend with density, we consider
  at each redshift two subsets of particles according to whether the
  density is above or below the virial overdensity $\Delta_{\rm vir}$
  (cf.  \Eq{eq:rvir} in Section~\ref{DMhalos}) and hence
  particles of the high-density subset either already belong to virialized structures
  or will be part of them at a later time.
  Table~\ref{tab:happi} gives
  the fraction of HAPPI particles in each subset.

 \begin{table}
   \caption{
     Fraction of HAPPI particles 
     in the subset of low-density particles, and in the subset of 
     high-density particles, respectively.}
 \label{tab:happi}
 \begin{tabular}{lcc}\hline
   redshift &  low-density & high-density  \\ 
   \hline \hline
   $z=0$ & 10\%      & 6\%    \\
   $z=0.5$ & 14\%      & 9\%    \\
   $z=1$ & 16\%      & 12\%    \\
   $z=4$ & 14\%      & 23\%    \\
 \end{tabular}
 \end{table}
 This raises the question about the exact locations as well as the
 "dynamics" of HAPPI particles.  A visual inspection shows that, at
 high redshift, they are preferentially located within the filaments
 flowing towards halos. At later times though, they can be found
 either in regions of strong dynamical activity (e.g., mergers, the
 outskirts of halos and infalling to halos), or at the very centres of
 relaxed systems. The smaller spatial force resolution of the \LCDM
 happi2 mentioned earlier can hence be linked to the influence of the
 HAPPI particles at the very centres of halos.

\section{Analysis of the matter distribution} \label{Analysis}

   \begin{figure*}
      \centerline{\psfig{file=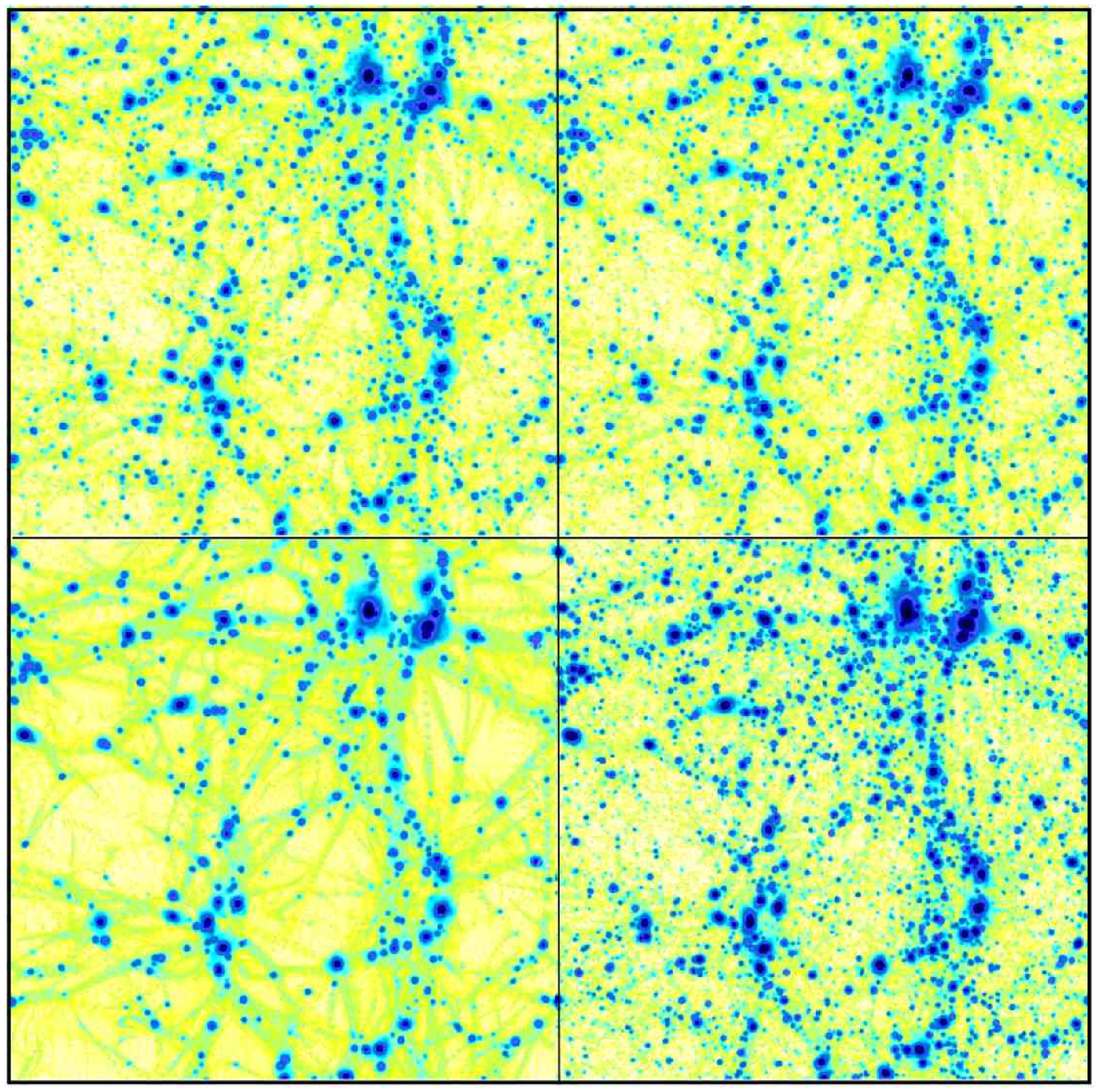,width=\hsize}}
      \caption{Color--coded density field of all four models
               at redshift $z=0$. The order is (clockwise starting
               upper left) \LCDM, \LCDM happi1, \LCDM happi2, 
               and \LWDM.}  
      \label{simu}
   \end{figure*}

   \begin{figure}
      \centerline{\psfig{file=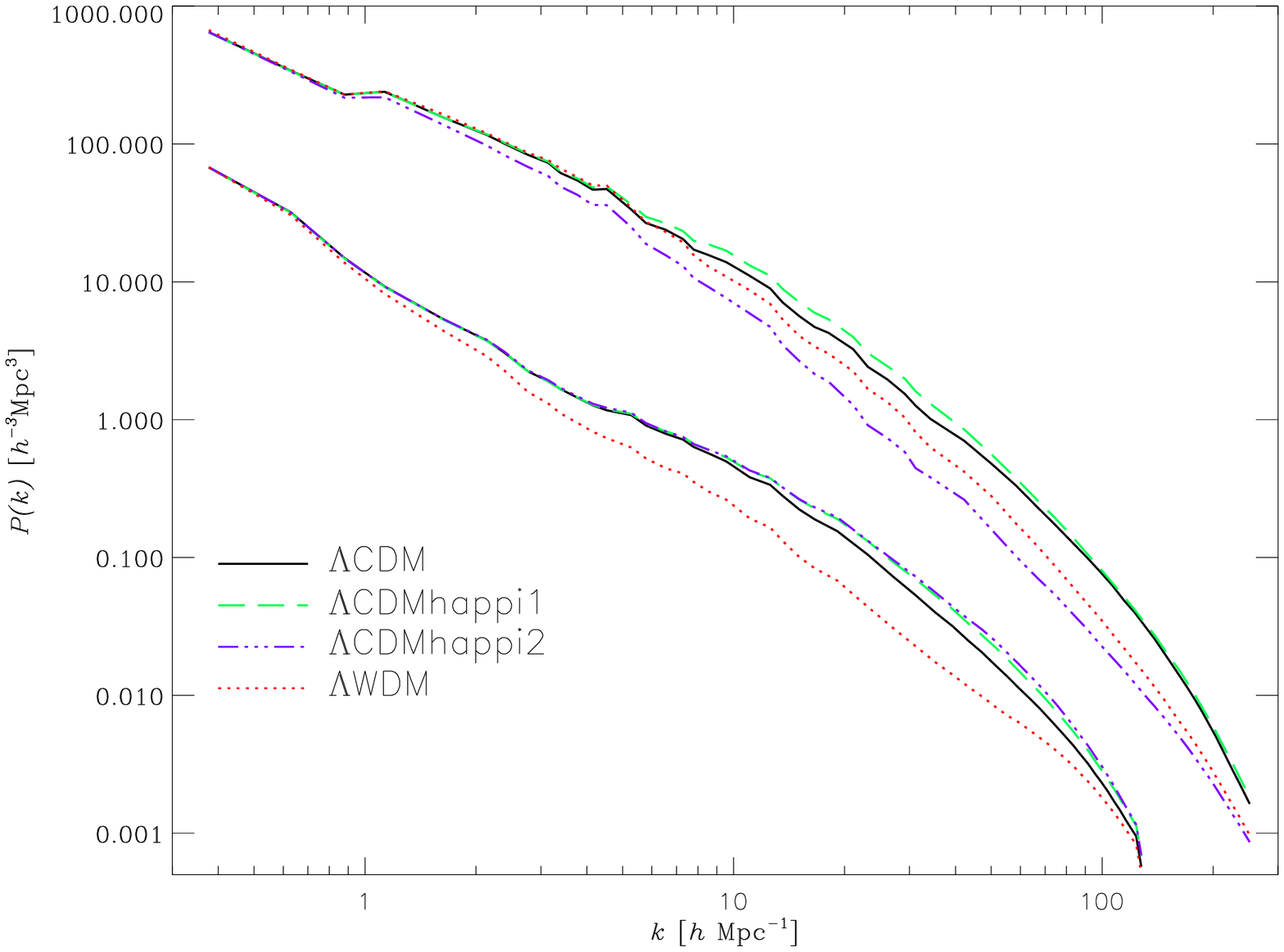,width=\hsize}}
      \caption{Dark matter power spectrum $P(k)$ for all models
               at redshifts $z=3$ (lower curves) and $z=0$ (upper curves).
             }  
      \label{power}
   \end{figure}

\subsection{Large Scale Structure and Global Properties}
A visual impression of the density field of the particle distribution
at redshift $z=0$ for all four models is shown in \Fig{simu}, where
the local density at each particle position was determined by
smoothing the distribution onto a regular grid (256$^3$ nodes) and
interpolating the density on the mesh back to the particle positions.
Not surprisingly, the \LWDM\ model appears less clumpy and far
smoother than the \LCDM\ simulation. However, there appears to be more
small scale structure in both the HAPPI runs, or at least the smaller
objects are more contrasted. We can readily relate this phenomenon to
the influence of $B$ and larger $B$ values give higher contrasts
(remember that the fiducial \LCDM\ model is nothing else than a HAPPI
model with $B$=0).  But despite the more grainy appearance of the
HAPPI runs in \Fig{simu}, the power on small scales is reduced
compared to the power of the \LCDM\ model. This can be verified in
\Fig{power}, where we plot the dark matter power spectrum of density
fluctuations for all four models at redshifts $z=3$ (lower curves) and
$z=0$ (upper curves).  Especially \LCDM happi2 falls behind \LCDM\ 
even though it appears to be marginally more evolved at higher
redshift. We will though see that these two results, the "graininess"
of the \LCDM happi2 model and the lack of small scale power, do not
exclude each other. The absence of small scale power on scales
below $\sim$ 1--2 \hMpc\ reflects the fact that the halos
corresponding to those scales (i.e. halos of mass $>10^{11}$\hMsun)
are internally less concentrated than their \LCDM\
counterparts.

\subsubsection{Beyond the two--point estimators}

Two--point estimators like the power spectrum are sensitive only to
the amplitude of the modes of the fluctuating field
$\varrho(\bx)-\varrho_b$. The information concerning the relative
phases is contained in the higher--order correlations. In the
literature there have been several meaningful quantities proposed
depending on higher--order correlations with a more or less
transparent physical interpretation. In this work, we have employed
the scalar {\em Minkowski functionals} (MFs) (see, e.g., Mecke \&
Stoyan 2000; Dom{\'\i}nguez 2001), which allow a quantification of the
morphological aspects appreciated by visual inspections. Given a density field
$\varrho(\bx)$ and a density threshold $\thr$, one constructs the
isodensity surface ${\cal S} = \{\bx | \varrho(\bx) = \thr\}$ (with
the convention that the region $\varrho>\thr$ is taken as the interior
of ${\cal S}$). The four MFs $V_{\nu}$, $\nu=0,1,2,3$, can be defined
as surface integrals over ${\cal S}$ and have the following
geometrical meaning (up to a conventional constant prefactor):
\begin{eqnarray*}
  V_0 & \propto & \textrm{volume enclosed by the isodensity surface 
    ${\cal S}$} \\ 
  V_1 & \propto & \textrm{total area of ${\cal S}$} \\
  V_2/V_1 & \propto & \textrm{mean curvature of ${\cal S}$, 
    averaged over ${\cal S}$} \\
  V_3/V_1 & \propto & \textrm{Gaussian curvature of ${\cal S}$, 
    averaged over ${\cal S}$}
\end{eqnarray*}
$V_0$ is in fact proportional to the probability that $\varrho(\bx) >
\thr$ (assuming spatial ergodicity of the realization). The ratio
$V_1/V_0$ is a measure of how compact the volume enclosed by ${\cal
  S}$ is packed. $V_2/V_1$ is a measure of the convexity of the
surface ${\cal S}$, while $V_3$ is proportional to the Euler
characteristic or genus of the body defined by ${\cal S}$:
\begin{eqnarray*}
  V_3 & \propto & \textrm{number of disconnected objects + 
    number of holes} \\
  & & \textrm{\mbox{} $-$ number of tunnels}.
\end{eqnarray*}
There seems to be a close relation between the threshold value at
which $V_3$ vanishes and the percolation threshold of the volume
enclosed by ${\cal S}$
(Mecke \& Wagner 1991; Neher 2003) --- as a matter of fact, the use of
percolation analysis is not rare in the analysis of cosmological
structures (e.g.  Yess \& Shandarin 1996). As an example of how the MFs
are to be interpreted, in App.~\ref{sec:poissonMF} we discuss the case
that $\varrho(\bx)$ is derived from a realization of a Poissonian
distribution of points.

\begin{figure}
  \centerline{\psfig{file=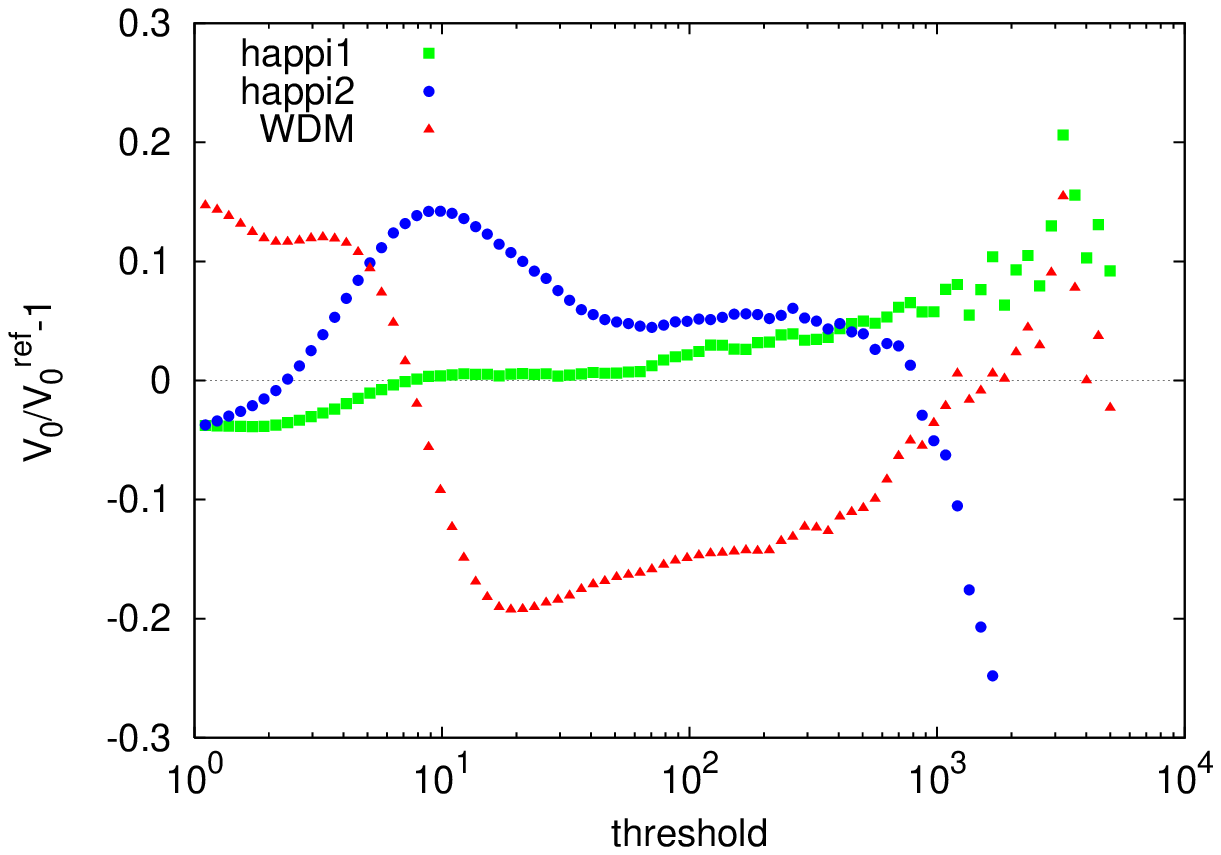,width=.95\hsize}}
  \centerline{\psfig{file=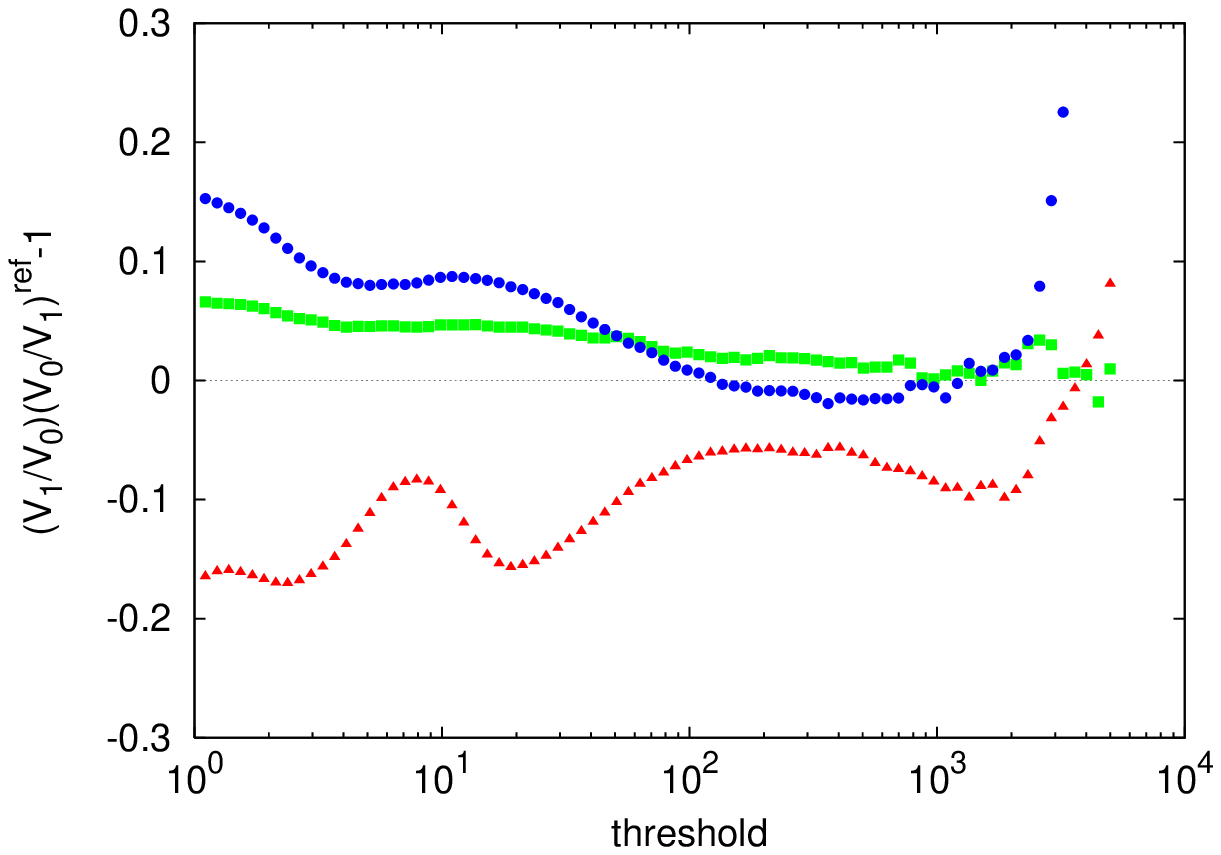,width=.95\hsize}}
  \centerline{\psfig{file=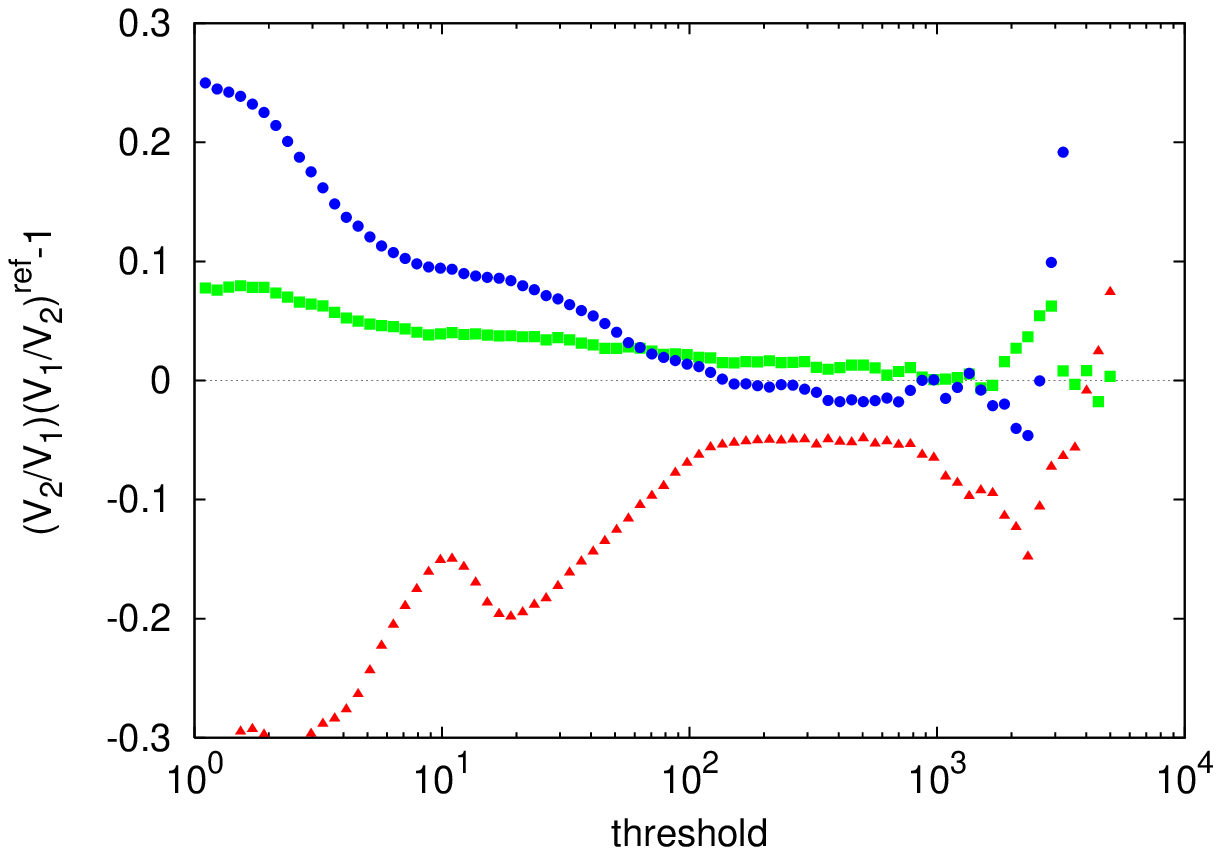,width=.95\hsize}}
  \centerline{\psfig{file=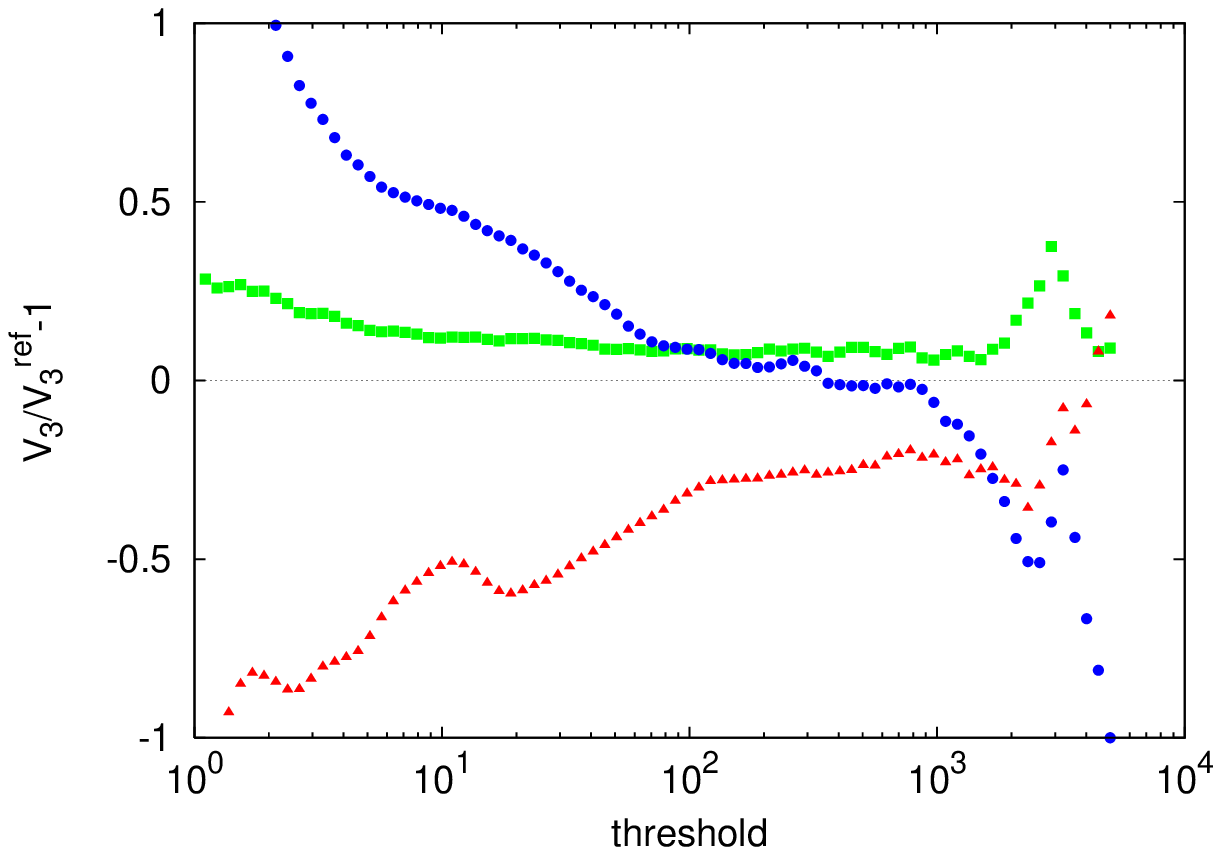,width=.95\hsize}}
  \caption{
    Minkowski functionals vs.~density threshold at redshift $z=0$. The
    threshold is given in units of particles per grid node. The grid
    has $128^3$ nodes, corresponding to a grid constant $0.20$\hMpc,
    so that the threshold coincides with the ratio
    $\varrho/\varrho_b$ for this spatial resolution.}
  \label{fig:MFs}
\end{figure}

Starting from the positions of the particles given by the simulation,
a density field $\varrho(\bx)$ is generated on a grid by smoothing
with the 
window~\Eq{TSC}. We generated 20 different realizations of the field
by displacing randomly the grid and we present the MFs averaged over
these realizations.  The MFs are functions of the grid constant and
the density threshold.  We observe that the measurement of the MFs in
an $N$-body simulation is affected by two kinds of errors (which are
more
pronounced for higher orders $\nu$ of the MF $V_\nu$): \\
(i) {\em Finite--volume effects}: the tails of the probability
distribution of the density field cannot be probed correctly in a
finite volume, so that the measured MFs are noisy and take discrete
values (due to the grid) as the threshold approaches the extremal
values probed in a given simulation. This effect can be reduced by
increasing the number of realizations. \\
(ii) {\em Finite--mass effects}: at threshold values corresponding to
$\lesssim$~1-10 particles per grid node, the MFs detect that the
density field is actually derived from a distribution of point
particles.

In order to emphasize the differences between the four simulations and
to facilitate the physical interpretation, we plotted as a function of
$\thr$ the relative difference between a measure in one model and the
corresponding one in the reference \LCDM\ model (denoted by 'ref'):
\begin{eqnarray*}
  \frac{V_\nu}{V_\nu^{\rm ref}}-1 &
  (\nu=0,3) , \\
  \frac{V_\nu}{V_{\nu-1}} \times
  \left(\frac{V_{\nu-1}}{V_\nu}\right)^{\rm ref} -1 &
  (\nu=1,2) .
\end{eqnarray*}
The plots are shown in \Fig{fig:MFs} for a grid of $128^3$ nodes. As a
general feature, the trend of the \LWDM\ model with respect to the
reference \LCDM\ model is always opposite to the trend of the two
HAPPI runs, with the deviations of \LCDM happi2 being larger than
those of \LCDM happi1. Only at very high density thresholds, \LCDM
happi2 seems to follow an opposite trend to \LCDM happi1: this is
because the maximum density in the \LCDM happi2 run is smaller than in
the \LCDM happi1 or \LCDM\ runs. This (physical) effect has
  been already noticed concerning the final force resolution of the
  runs: it is due to the larger effect of the additional acceleration
  ${\bf C}$ and can be directly linked to the location of "HAPPI
  particles" at redshift $z=0$ at the center of relaxed structures,
  see Secs.~\ref{setup} and \ref{HAPPIimportance}.

%
The results can be summarized as follows:
\begin{itemize}
  
\item The dependence $V_0(\thr)$ quantifies the visual impression that
  low--density regions ($\varrho \lesssim 10 \varrho_b$) are clearly
  less likely in the WDM model, while overdense regions ($10 \lesssim
  \varrho/\varrho_b \lesssim 10^3$) are more abundant in the HAPPI
  models. As remarked, higher density 
  areas are rarer in \LCDM happi2 and more frequent in \LCDM happi1
  compared to \LCDM.
  
\item From the fact that $V_2, V_3 >0$ (not shown in the plot) in the
  whole range of threshold values, one infers that, observed at the
  resolution of $128^3$ nodes, the matter distribution consists mainly
  of ``clusters'' (i.e., disconnected objects which are convex on
  average). According to the plot of $(V_2/V_1)(\thr)$, these clusters
  in the HAPPI runs tend to be rounder than in the \LCDM\ run,
  while they tend to be more cigar--shaped (filament--like) in the
  \LWDM\ model. From the plot of $V_3(\thr)$ one deduces that the
  \LWDM\ model has less clusters at all threshold values; the HAPPI
  runs have more clusters, except for \LCDM happi2 at very high
  densities because this particular model does not reach as high densities as the
  other ones.
  
\item Finally, the plot of $(V_1/V_0) (\thr)$ shows that matter is
  more compactly packed in the \LWDM\ model, and less compactly in the
  HAPPI runs. This is likely due to the different cluster abundances
  just discussed.

\end{itemize}
We have repeated the analysis at different spatial resolutions
($16^3$, $32^3$, $64^3$ and $256^3$ nodes) for the Minkowski functionals. The quantitative
differences between models decrease as the grid becomes coarser, but
the same conclusions hold roughly in a qualitative manner.

Summarizing, the \LWDM\ run has less voids and more filamentary--like
structures than the reference \LCDM\ model, while the HAPPI runs have
comparatively more mass concentrated in small, roundish clusters.

\subsubsection{The velocity field}

Motivated by the theoretical discussion, we have also addressed the
distribution of the (comoving) vorticity, $\bomega = \nabla \times
\bu$, and the (comoving) divergence, $\theta = \nabla \cdot \bu$, of
the peculiar velocity field $\bu(\bx)$. Using the positions and
velocities of the particles, we compute the fields $\bomega(\bx)$ and
$\theta(\bx)$ on a regular grid (see Sec.~\ref{Nbody}). The cumulative
probabilities that $|\bomega|^2 > \hat{\omega}^2$ and that $\theta^2 >
\hat{\theta}^2$ are given by the first Minkowski functional $V_0$,
which we compute as explained previously.

\begin{figure}
  \centerline{\psfig{file=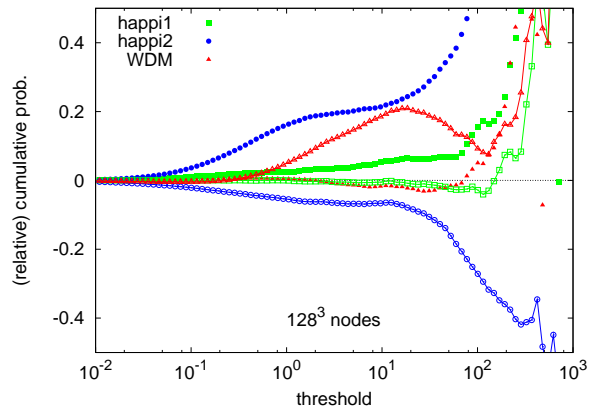,width=.95\hsize}}
  \caption{ Cumulative probabilities $P(|\bomega|^2 >
    \hat{\omega}^2)$, $P(\theta^2 > \hat{\theta}^2)$ relative to the
    probabilities in the \LCDM\ model at redshift $z=0$ at a
    resolution of $128^3$ grid nodes. The solid symbols represent the
    (relative) cumulative probability of the vorticity, the joined
    open symbols correspond to the divergence. The threshold
    ($\hat{\bomega}^2$, $\hat{\theta}^2$) is given in units of
    $H_0^2$.}
  \label{fig:vortdiv}
\end{figure}

\Fig{fig:vortdiv} shows $P/P^{\rm ref}-1$, where $P$ is the cumulative
probability measured in a simulation, and $P^{\rm ref}$ is the
corresponding probability in the reference \LCDM\ simulation. The
\LCDM happi2 model shows a clear tendency to have a larger vorticity
and a lower divergence (in absolute value) than the \LCDM\ model. The
vorticity in the \LCDM happi1 model exhibits a similar tendency, but
the differences are quantitatively smaller. 
%
In order to minimize finite--mass effects, we considered only grid
nodes for which the value of the smoothed density field corresponds to
more than 10 particle per node. If we restrict the measurement of the
probability distributions to higher densities, the quantitative
differences tend to be smaller.

\subsection{Dark Matter Halos}\label{DMhalos}

The analysis in the following Subsection is primarily based upon
gravitational bound objects which were identified using \mlapm's Halo
Finder (\mhf) (Gill, Knebe~\& Gibson 2004). This newly developed halo
finder uses the adaptive grid structure invoked by the \nbody\ code
\mlapm\ and re-organizes it into a tree. The centres of the grids at
the end-leaves of a branch of the tree serve as (potential) halo
centres and all gravitationally bound particles about these centres
are being collected. For a more elaborate discussion of this halo
finder we refer the reader to Gill~\ea (2004). We only like to note at
this point that \mhf\ is essentially parameter free and naturally
finds halos with exactly the same accuracy as the simulation.
Besides of the growth of the halo mass function and the abundance
evolution of gravitationally bound objects, respectively, we confine our analysis to objects identified
at redshift $z=0$. The investigation of the evolutionary history and
hierarchical growth of structures will be published elsewhere.

\subsubsection{Global Properties}
\begin{figure}
      \centerline{\psfig{file=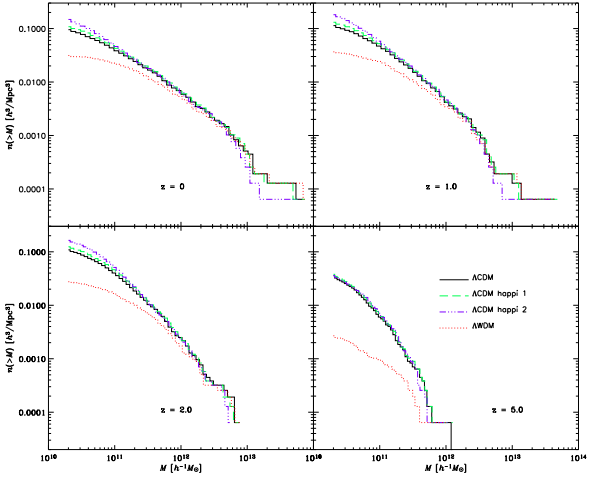,width=\hsize}}
      \caption{Cumulative mass function $n(>M)$ of DM halos. 
      }
      \label{massfunc}
   \end{figure}

\begin{figure}
      \centerline{\psfig{file=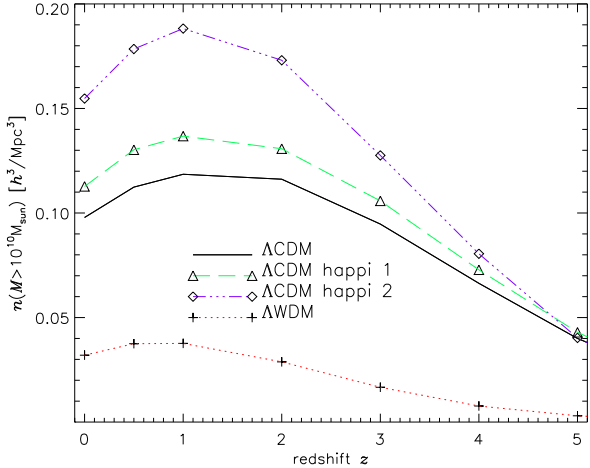,width=\hsize}}
      \caption{Number density evolution of objects more massive than 
        $M>10^{10}$\hMsun.
      }
      \label{abundance}
   \end{figure}

In \Fig{massfunc} we show the evolution of the cumulative mass
function $n(>M)$ of dark matter halos, i.e., the number of DM halos
with mass larger than $M$. The \LWDM\ model clearly has less low--mass
objects as already pointed out by other authors (Knebe~\ea 2002, Bode~\ea 2001, Col{\'\i}n~\ea 2000). All three CDM models though perfectly agree at a redshift of $z=5$,
but there is a clear trend for the HAPPI models to give rise to more
small mass halos, in agreement with the conclusion derived in the
previous Section. This can be verified in \Fig{abundance}, where
we plot the number density evolution of objects more massive than
$M>10^{10}$\hMsun\ ($>$ 20 particles). In \LCDM happi2 there are
roughly 50\% more halos above our mass cut than in the reference
\LCDM\ run.

   \begin{figure}
      \centerline{\psfig{file=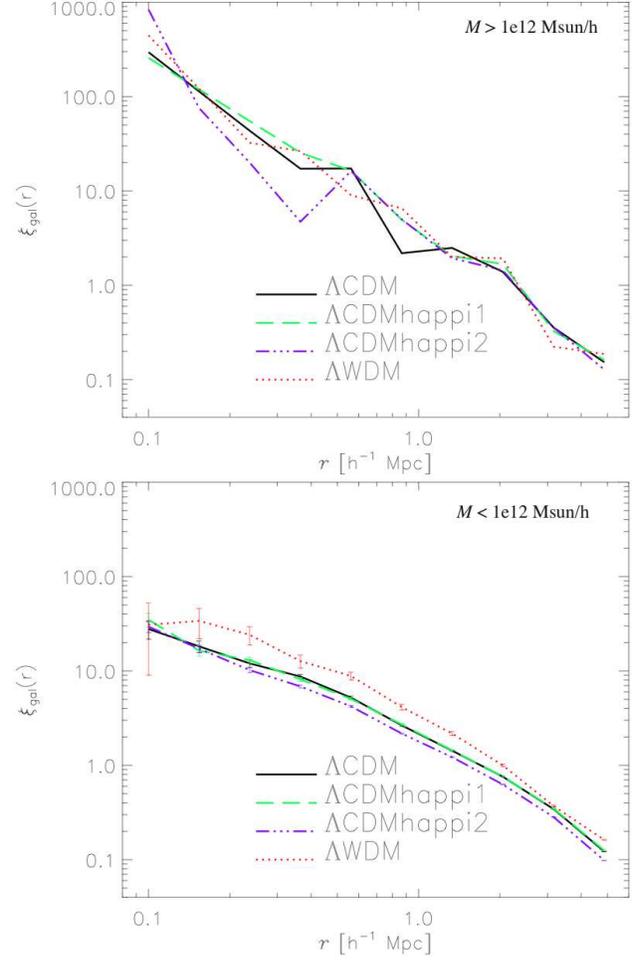,width=\hsize}}
      \caption{Two-point correlation function of objects more 
        massive (upper panel) and less massive (lower panel) than
        $10^{12}$\hMsun.  
      }
             \label{halohaloXi}
   \end{figure}

In order to study the clustering properties of these halos we estimated the
two--point correlation function for low-- and high--mass objects, respectively, as
\begin{equation}
  \xi_{\rm gal}(r) = -1 + \frac{V}{N_{\rm gal}^2} 
  \sum_{\alpha=1}^{N_{\rm gal}} 
  \frac{n_\alpha(r; \Delta r)}{v(r; \Delta r)} ,
\end{equation}
where $N_{\rm gal}$ is the total number of objects in the simulation
volume $V$, and $n_\alpha(r; \Delta r)$ is the total number of objects
in a spherical shell of radius r and thickness $\Delta r$ (and volume
$v(r; \Delta r)$), centered at the $\alpha$-th object.
The result (along with Poisson error bars based upon the
  number of pairs in each bin) is shown in \Fig{halohaloXi}.
The low--mass objects show very similar clustering patterns, with a higher amplitude of $\xi_{\rm gal}$ for \LWDM\ 
and a marginally decreased correlation for the
\LCDM happi2 model.
The situation though is difficult to judge at the high--mass end as we
have too few pairs per bin to make conclusive statements. The errors
bars are larger than the differences amongst the models and hence have
been omitted. It seems, however, that the clustering pattern of high
mass objects is similar in all models and does not show differences
when including the HAPPI correction term.

The most striking (and interesting) difference between standard \LCDM\ and the 
two HAPPI models, however, emerges when we turn to the spin
parameter distribution. The spin parameter $\lambda$ was calculated
using the definition given by Bullock~\ea (2001a),

\begin{equation} \label{lambda}
 \lambda = \frac{|{\bf L}|}{\sqrt{2} M_{\rm vir} v_{\rm vir} r_{\rm vir}} \ ,
\end{equation}

\noindent
where ${\bf L}$ is the angular momentum of the halo with
  respect to its center of mass, $r_{\rm vir}$ is the virial radius
of the halo, $M_{\rm vir}$ is the virial mass (mass enclosed within
the virial radius), and $v_{\rm vir} = \sqrt{G M_{\rm vir}/r_{\rm
    vir}}$ is the circular velocity at the virial radius. The virial radius and mass are
determined by the condition
\begin{equation}
  \label{eq:rvir} 
  M_{\rm vir} = \frac{4 \pi}{3} r_{\rm vir}^3 \varrho_{\rm vir} ,
\end{equation}
where $\varrho_{\rm vir} = \Delta_{\rm vir} \varrho_b(z=0)$ is a fiducial
density with $\Delta_{\rm vir}\approx 340$ (at redshift $z=0$) based on the dissipationless
  spherical top-hat collapse model for the cosmological parameters of
  the \LCDM\ model.
%
The probability distribution, $P(\lambda)$, of the spin parameter was
fitted to a log--normal distribution (e.g. Frenk~\ea 1988; Warren~\ea
1992; Cole~\& Lacey 1996; Maller, Dekel~\& Somerville 2002; Gardner
2001),

\begin{equation} \label{lognormal}
 P(\lambda) = \displaystyle \frac{1}{\lambda \sqrt{2\pi\sigma_0^2}}
              \exp \left( {-\frac{\ln^2 (\lambda/\lambda_0)}{2 \sigma_0^2}} \right) \ .
\end{equation}

\begin{figure}
      \centerline{\psfig{file=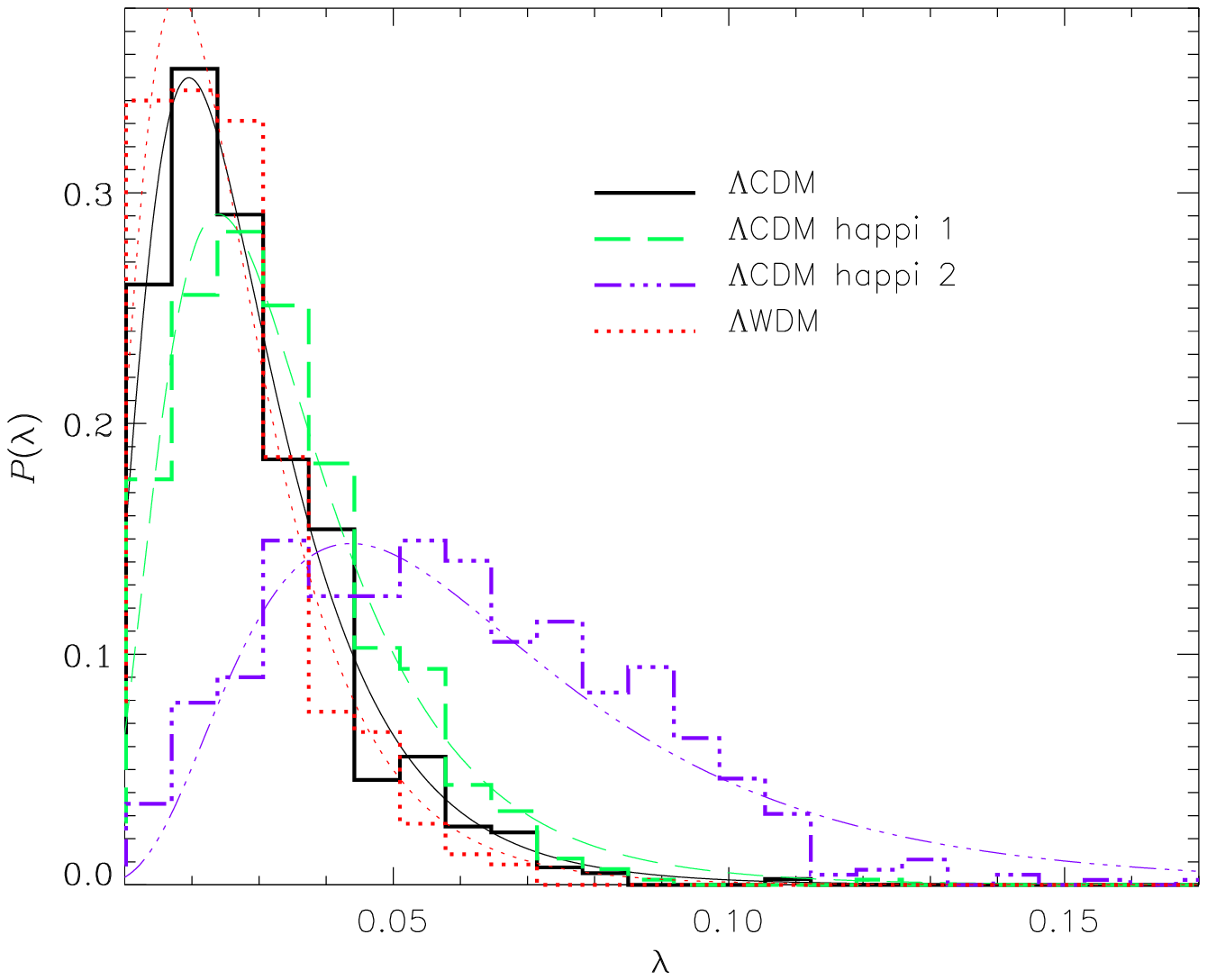,width=\hsize}}
      \centerline{\psfig{file=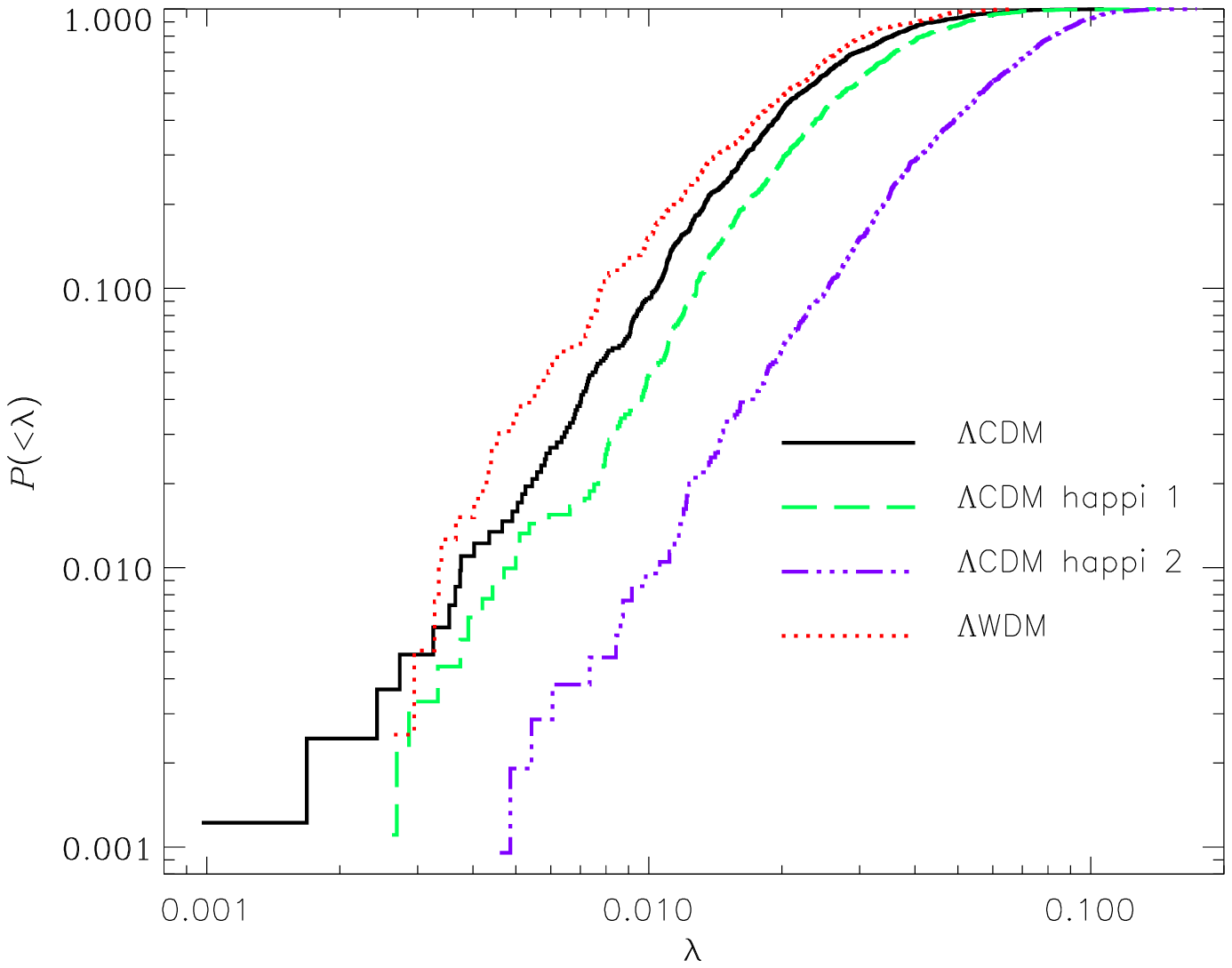,width=\hsize}}
      \caption{
        Upper plot: Measured spin parameter distribution for all four
        models at redshift $z=0$, and the corresponding fit to
        \Eq{lognormal}.
        Lower plot: Measured cumulative distribution of the spin
        parameter.
      }
      \label{SpinFit}
   \end{figure}

\noindent
The results are presented in \Fig{SpinFit} and in
Table~\ref{SpinParam}, showing that a larger value of $B$
entails a larger spin parameter. 
In the lower panel of \Fig{SpinFit} -- where we plot the cumulative
distribution of the spin parameter -- we clearly see that for a given
$\lambda$ the probability to find halos with a \textit{smaller}
$\lambda$ is greatly lowered in \LCDM happi2 --- or in other words, it is more
likely to find halos with larger spin parameters in \LCDM happi2. 
%
%
As we will see later on (cf.\ \Fig{JJ} in Section~\ref{SECCrossCorr}),
there is a mass dependence of this result: lower mass halos tend to
dominate the signal seen in \Fig{SpinFit}.  When plotting the
mass--weighted spin parameter distribution (not shown)
the peaks for all four models approach each other. However, even then the HAPPI runs 
still show a distinct tail out to larger $\lambda$ values.

\begin{table}
\caption{
  Parameters derived from fitting the spin parameter distributions
  $P(\lambda)$ to \Eq{lognormal}.}
\label{SpinParam}
\begin{tabular}{lcc}\hline
 model        &  $\lambda_0$ & $\sigma_0$  \\ 
\hline \hline
 \LCDM        &   0.0287      &   0.4485    \\
 \LCDM happi1 &   0.0333      &   0.4611    \\
 \LCDM happi2 &   0.0596      &   0.4914    \\
 \LWDM        &   0.0259      &   0.4001    \\
\end{tabular}
\end{table}

The HAPPI correction term also affects the
concentration of dark matter halos. 
We define the concentration \c15 as the
ratio of the virial radius \rvir\ and the radius of the sphere
that contains $1/5$ of the virial mass (i.e. $r_{1/5}$ is defined
via $M(<r_{1/5}) = M_{\rm vir}/5$):

\begin{equation} \label{c15}
  c_{1/5} = \frac{r_{\rm vir}}{r_{1/5}} \ .
\end{equation}

\begin{figure}
      \centerline{\psfig{file=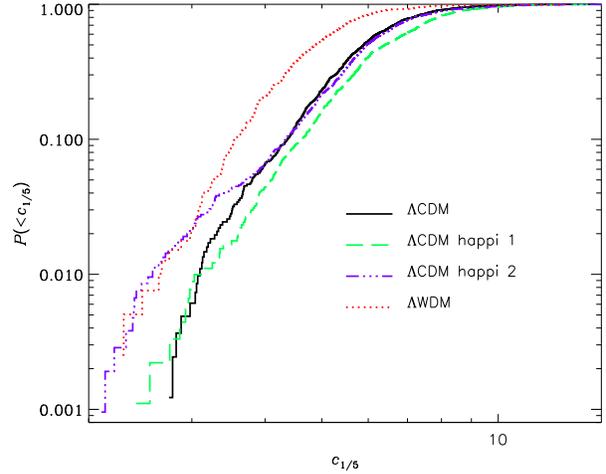,width=\hsize}}
      \caption{Cumulative distribution of the concentration parameter 
        \c15.
             }  
      \label{PCconc}
   \end{figure}

\noindent
In view of the definition~(\ref{eq:rvir}), it follows that the average
density within a radius $r_{1/5}$ is given by $(1/5) c_{1/5}^3
\varrho_{\rm vir}$.
\Fig{PCconc} plots the cumulative probability distribution of the
concentration of halos.
We observe an obvious trend for an overabundance of low--concentration 
halos in the \LWDM\ and \LCDM happi2 models. However, the opposite actually holds for
\LCDM happi1, where there appear to be of order 10\% more concentrated
halos.
%
The relative lack of power on scales $\lesssim 1$\hMpc\ noted in
\Fig{power} for WDM and \LCDM happi2 is related to the relatively 
lower concentration (and increased smoothness for WDM) 
of the halos observed in these models (one must bear in
mind that the halos have a virial radius $\lesssim 800$\hkpc, see
\Fig{DensProfile}).

Although not shown, we confirm that the scaling of the concentration
$c_{\rm 1/5}$, defined by \Eq{c15}, with mass $M_{\rm vir}$ follows
the same relation as the one proposed by Bullock~\ea (2001b) for the
"NFW concentration" of the halo,
$c_{\rm NFW}=r_{\rm vir}/r_s$ (see \Eq{NFWfit} for the definition of $r_s$).

\subsubsection{Cross Correlations} \label{SECCrossCorr}
While the last Section dealt with the distribution of halo properties,
we now compare these properties across the models, i.e.
how do these properties change in a given halo when moving from one
model to another?

\begin{figure}
      \centerline{\psfig{file=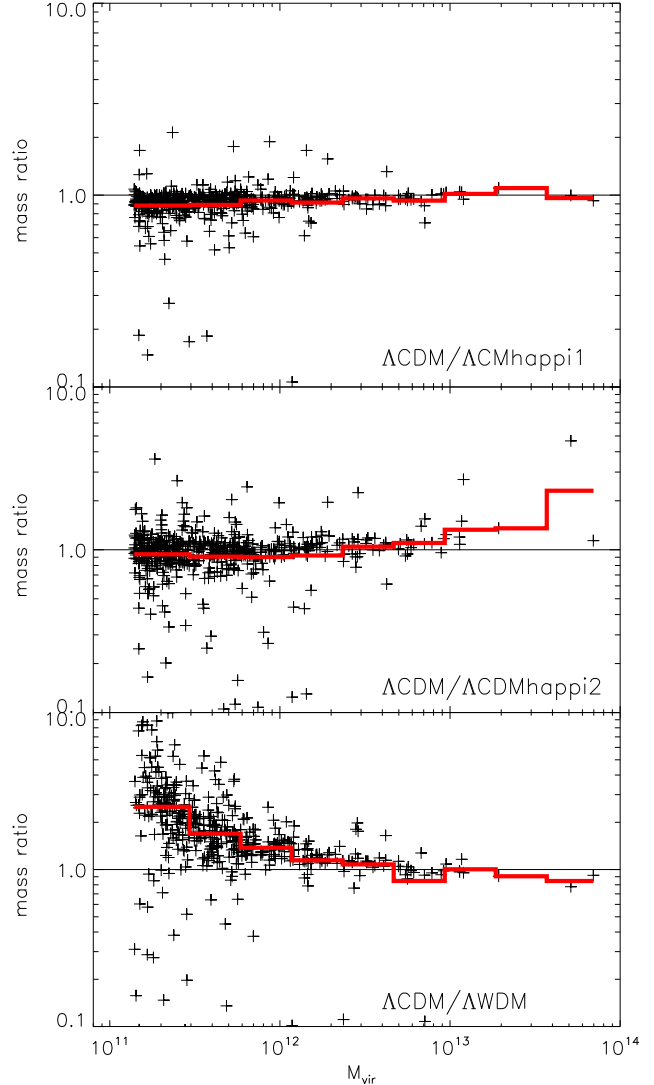,width=\hsize}}
      \caption{Ratio of halo masses for cross-identified objetcs.
        The virial mass $M_{\rm vir}$ is given in units of
          \hMsun. The histograms represent the mean ratio in the
        respective bin.
             }  
      \label{MassMass}
   \end{figure}

\begin{figure}
     \centerline{\psfig{file=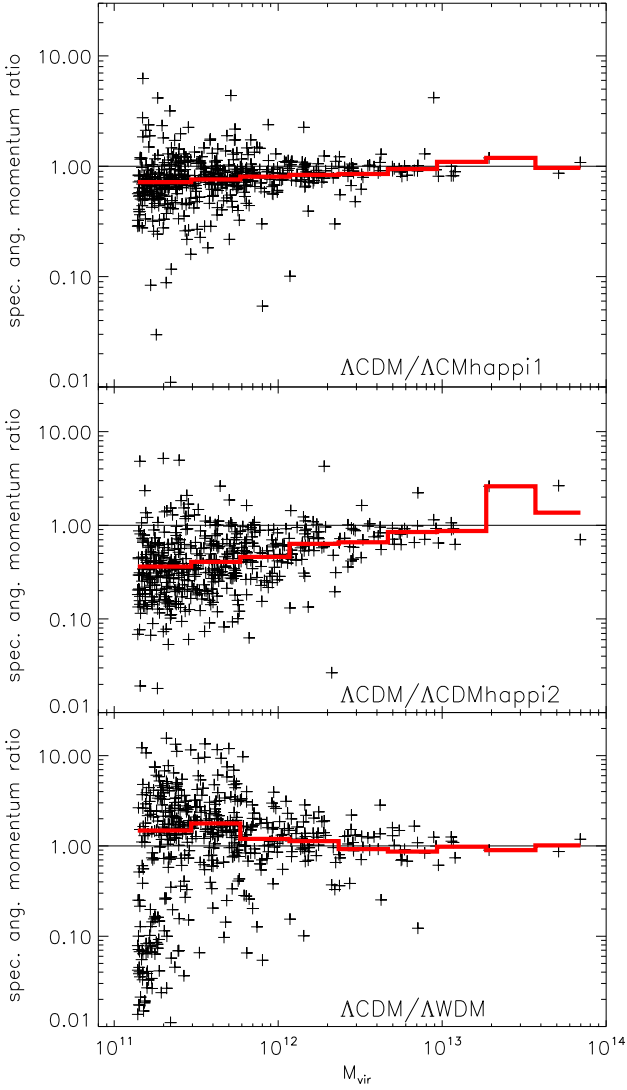,width=\hsize}}
      \caption{Ratio of specific angular momenta. 
      }  
      \label{JJ}
   \end{figure}

\begin{figure}
      \centerline{\psfig{file=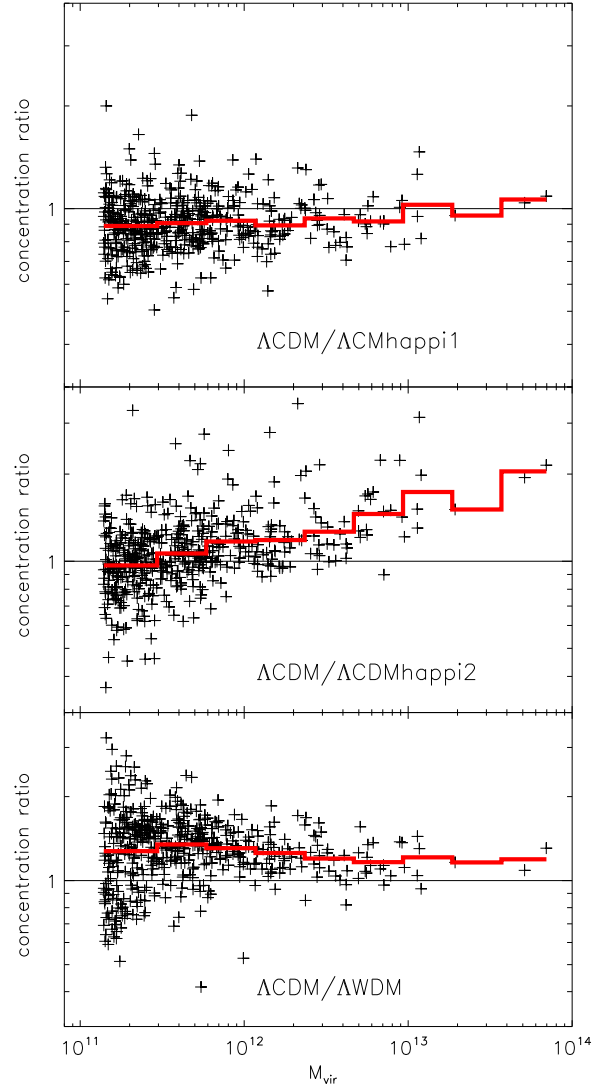,width=\hsize}}
      \caption{Ratio of halo concentrations \c15. 
      }  
      \label{ConcConc}
   \end{figure}

In order to find corresponding halos across the four models, we
compare their individual particle content.  We start with a halo in
the \LCDM\ model, whose particles are tagged, and locate the
corresponding halo in the other three models as the one that shares
the largest number of tagged particles.  In the
Figs.~(\ref{MassMass}--\ref{ConcConc}) 
we always plot the value of the property under investigation in the
\LCDM\ model, divided by the value of said property in the other model
for all "cross--identified" halos.  These "scatter-plots" are always
accompanied by histograms, where we average the ratios presented in
the respective figure in nine bins across the actual mass range.  The
percentages of counterparts amount to practically 100\% for the HAPPI
models while there are only 40\% cross-identified halos in the \LWDM\
model. This number though increases to about 90\% when we consider
halos containing more than 200 particles (i.e.,
  $M \grtsim 10^{11}$\hMsun), and hence we set this as a lower limit in the
  cross--correlation plots.
%

We start with the most obvious halo attribute, namely the halo mass itself. 
\Fig{MassMass} shows that there is a very tight correlation for the
masses of individual halos, especially for \LCDM\ and \LCDM happi1.
The scatter about the 1:1 relation (the flat line of value 1)
marginally increases from $\sim$11\% (averaged $1\sigma$-value) for
\LCDM happi1 to $\sim$18\% for \LCDM happi2.  At the
  high--mass end, the \LCDM happi2 halos tend to have a slightly larger
  mass.  The most pronounced differences can be found for \LWDM\ 
though: at the low--mass end the halos in \LWDM\ appear to be less
massive than their \LCDM\ counter parts.

In the previous Section we showed that the HAPPI correction term
lead to an increase in angular momentum by investigating the
spin parameter distributions. But can we be sure that the observed
rise in $\lambda$ as defined by \Eq{lambda} is not related
to a possible decrease of virial radius \rvir\ and/or virial velocity
$v_{\rm vir}$? To clarify this uncertainty we show in \Fig{JJ} the cross 
correlation of total specific angular momentum, 

\begin{equation}
 J = \frac{|{\bf L}|}{M_{\rm vir}} \ .
\end{equation}

\noindent
In view of the already mentioned minimal scatter in the mass of
cross-identified halos between the \LCDM\ model and the HAPPI models,
\Fig{JJ} confirms the previous result that one effect of the term ${\bf C}$ in
\Eq{sse_momentum} is to inject angular momentum to
halos.
We also note that there is a mass dependence in this trend: the
differences in angular momentum are on average larger for lower mass
objects, this being particularly noticeable for the \LCDM happi2 and
\LWDM\ models; this "break" roughly happens at around $10^{12}$\hMsun.

We close this Section with an investigation of the cross-correlation of the concentration parameter
\c15\ in \Fig{ConcConc}. The results are consistent with \Fig{PCconc}: \LCDM happi1 halos are more concentrated than their \LCDM\ counterparts, while the 
excess of low--concentration halos for \LCDM happi2 is due to
high--mass halos. 
The mass trend already noted in \Fig{JJ} can also be acknowledged in this figure.

Finally, we mention as a general property that the dispersion
  in the scatter plots increases as the halo mass diminishes. 
We attribute most of this scatter
  to differences in the halo's formation history,
  but a detailed study as function of redshift is required
  which we will postpone to a later paper.
  Moreover, numerical effects could also contribute to some
  extent.

\subsubsection{A Closer View of Individual Halos}
\begin{figure}
      \centerline{\psfig{file=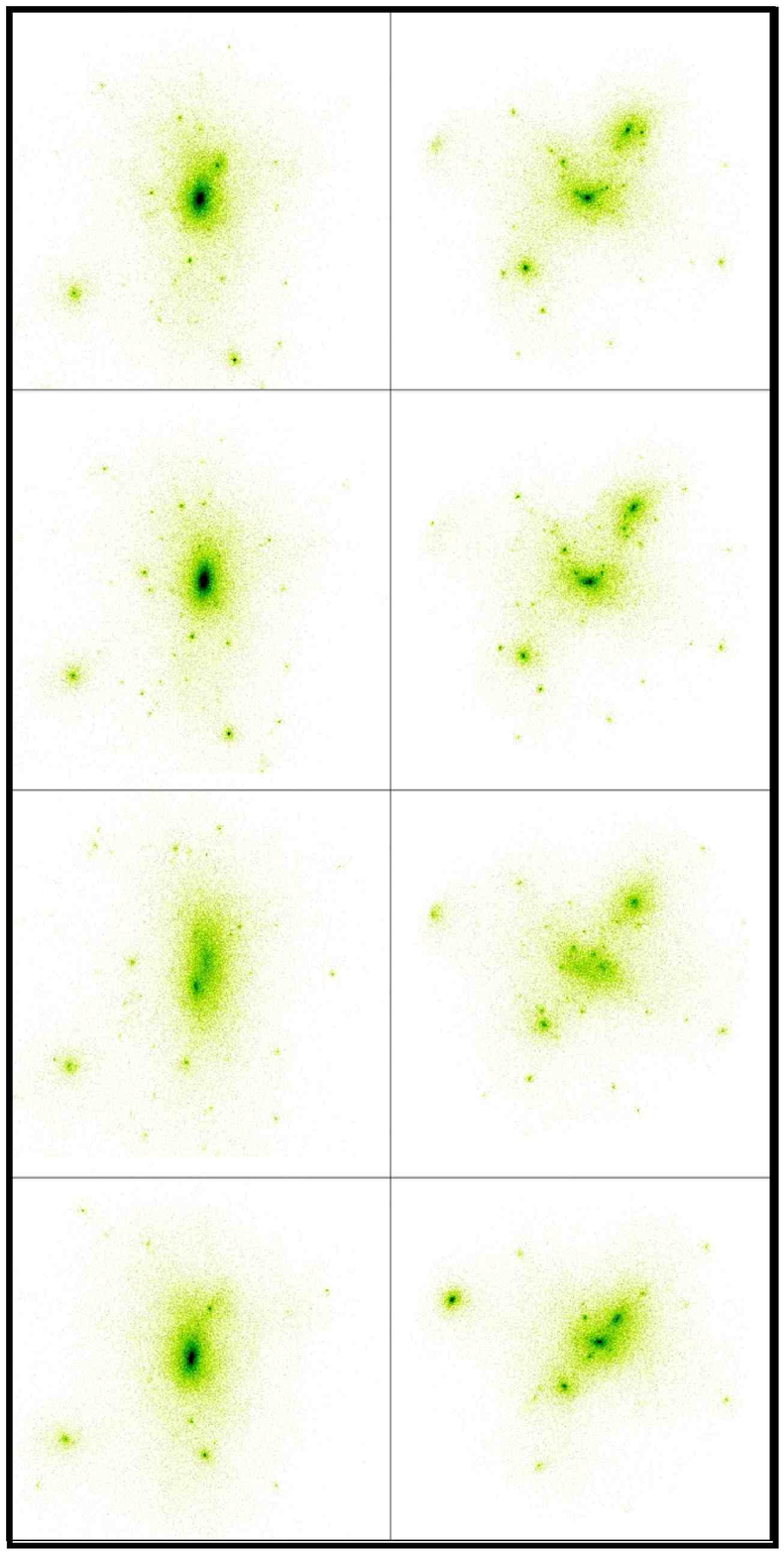,width=\hsize}}
      \caption{Visual representation of the two most massive halos
               at redshift $z=0$. The left column shows the
               most massive halo while the right column is
               the second most massive one.
               From top to bottom one has \LCDM, \LCDM happi1, 
               \LCDM happi2, and \LWDM.}  
      \label{Halos}
   \end{figure}

A visual representation of the two most massive halos in all four models
is given in \Fig{Halos}. This figure nicely demonstrates the result regarding the lower concentrations
in (high mass) \LCDM happi2 halos:
the second most massive halo does not even show a distinct centre in \LCDM happi2
and appears more "puffy" than in any of the other models. 

The question now arises whether the density profiles of the HAPPI models
can still be fitted by the universal density profile advocated
by  Navarro, Frenk~\& White (NFW, 1997)

\begin{equation} \label{NFWfit}
 \rho(r) = \frac{\rho_s r_s^3}{r(r_s+r)^2} \ .
\end{equation}

\begin{figure*}
      \centerline{\psfig{file=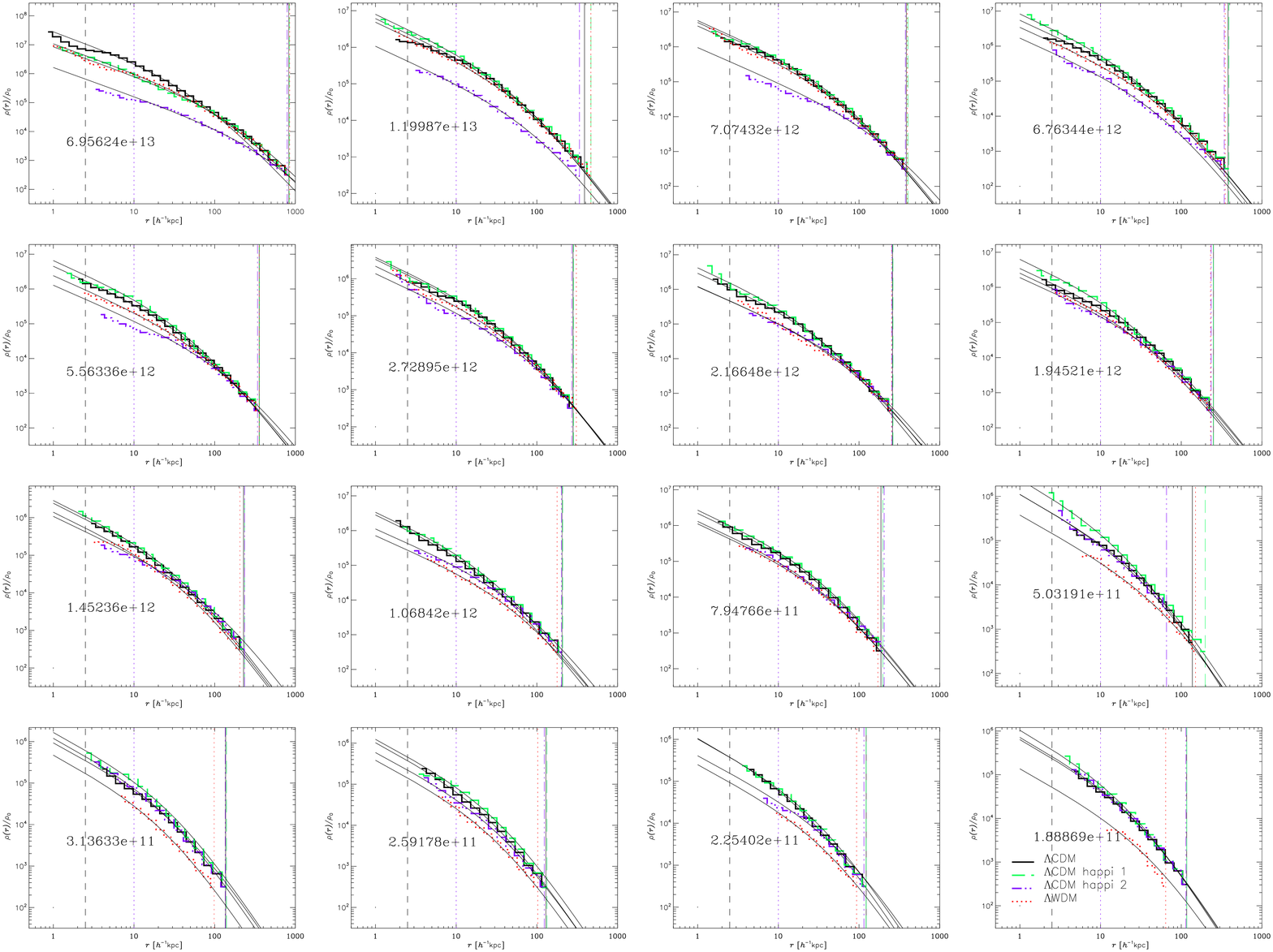,width=\hsize}}
      \caption{Density profiles for halos at redshift $z=0$. The number
        printed into each halo panel is the mass of the halo in units
        of \hMsun. 
        The vertical lines to the right indicate the respective virial
        radius while the two vertical lines to the left indicate the
        spatial resolution of the \LCDM, \LCDM happi1, \LWDM\ (dashed)
        and \LCDM happi2 (dotted) run, respectively.  }
      \label{DensProfile}
   \end{figure*}

\noindent
\Fig{DensProfile} now shows $\rho(r)$ and the
corresponding best fits to NFW profiles for a selection of halos
covering the mass range from the most massive one (upper left) to
rather light halos (lower right) containing a mere 300 particles.
For (nearly) all \LCDM happi2 halos we observe a relative
flattening in the central regions. 

In order to gauge the quality of the fits (for the 16 presented sample
profiles) we calculate the $\chi^2$ value defined as 

\begin{equation}\label{chi2NFW}
 \chi^2 = \frac{1}{N_{\rm bins}} 
          \sum_{i=1}^{N_{\rm bins}}
           \left|\frac{\rho_i - \rho_{\rm NFW}(r_i)}
                {\rho_{\rm NFW}(r_i)}\right|^2 ,
\end{equation}

\noindent
where $\rho_i$ are the binned density profiles derived from the
simulation data and $\rho_{\rm NFW}$ the best--fit NFW profiles.
This analysis then indicates that all four models are equally well
fit by \Eq{NFWfit} with $\chi^2$ varying in the range $\chi^2 \sim 0.02-0.05$
depending on the weighing scheme applied for each individual bin.
This entails that the dark matter halos of the HAPPI runs still exhibit the rather infamous "cusp" at the center.

We remind the reader again that the
force resolution throughout the runs varies. Whereas \LCDM, \LCDM happi1,
and \LWDM\ reached 2.5\hkpc\ resolution, \LCDM happi2 reliably
resolves structures only on scales larger than 10\hkpc. Moreover, the
resolution can also change from halo to halo due to the adaptive mesh
nature of both the halo finder and the \nbody\ code: not all halo centres
lie on the finest grid level reached in the simulation. However, we plot profiles
starting from the distance $r_{\rm min}$ that
corresponds to a sphere containing at least 10 particles (and hence $r_{\rm min}$ can be actually smaller than
the nominal resolution of the simulation).

\begin{figure*}
      \centerline{\psfig{file=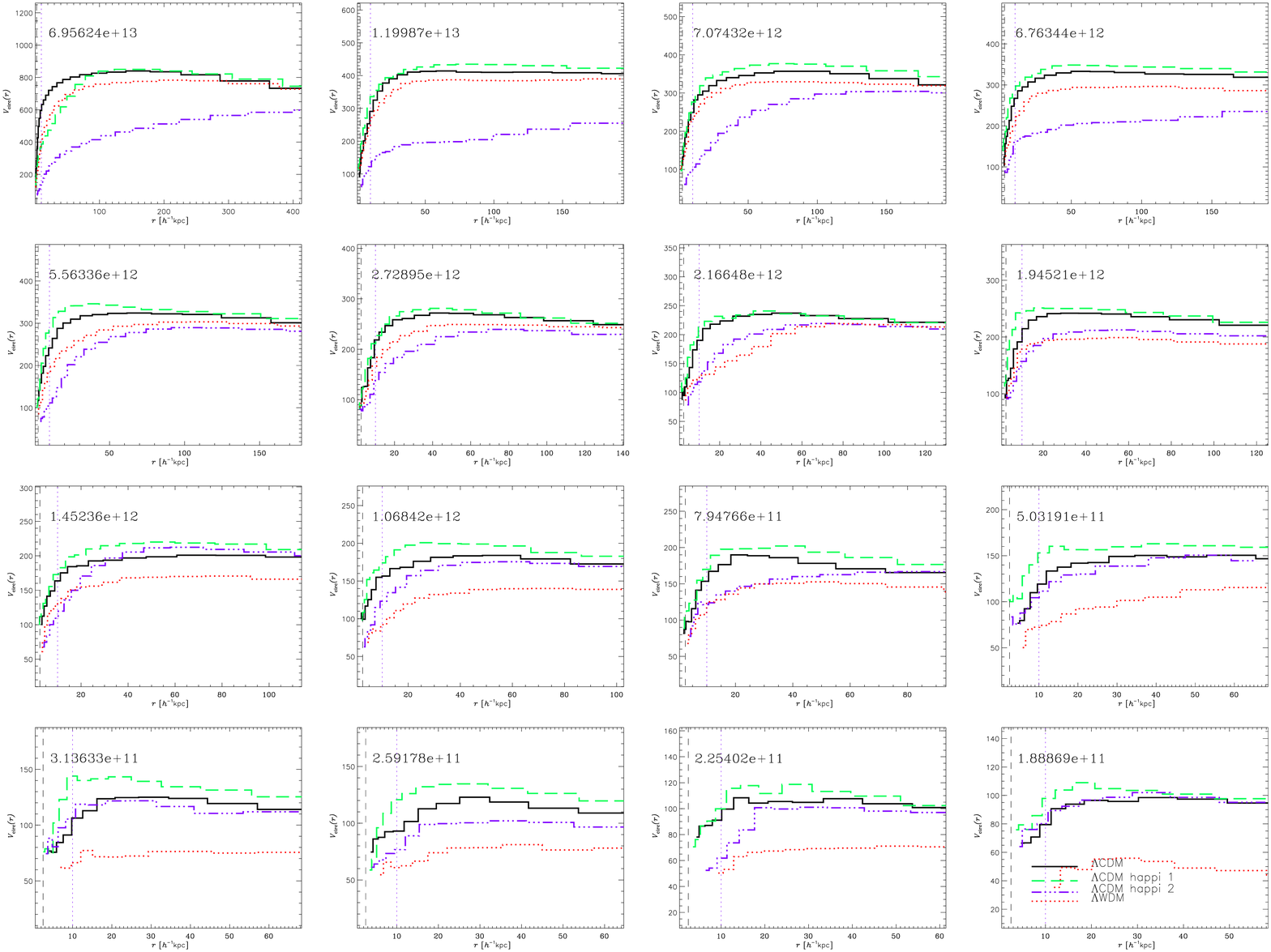,width=\hsize}}
      \caption{Rotation curves for the same halos as in \Fig{DensProfile}.
      }  
      \label{VcircProfile}
   \end{figure*}

For the same set of halos we present in \Fig{VcircProfile} the rotation curves
out to half the respective virial radius. 
The rotational velocity $v_{\rm circ}(r)$ is defined as

\begin{equation}
  \label{eq:vcirc}
  v_{\rm circ}(r)^2 = \frac{G M(<r)}{r} .
\end{equation}
There are a number of interesting observations to discuss now. We find that in nearly every halo 
the \LCDM happi1 rotation curve rises to higher values
than any of the other models. While the maximum is still at comparable
distances in \LCDM\ and \LCDM happi1, the latter shows a steeper
inner increase and a subsequent faster decline to nearly the same
level in the "outer" parts. Moreover, the \LCDM happi1 rotation
curves are \textit{always} slightly above the corresponding \LCDM\ 
curves.
Quite the opposite is true for \LCDM happi2. Here we find that in most
of the cases the circular rotation values at a given radius
are substantially smaller than in \LCDM. However, this difference
becomes less prominent in lower mass systems and the flat part of the
rotation curve nearly reaches the same level as \LCDM.
This discrepancy in the trends between \LCDM happi1 and \LCDM happi2 for
high--mass halos is likely related to the also opposite trends
concerning the concentration, Figs.~\ref{PCconc}~and~\ref{ConcConc},
and the small--scale power, Fig.~\ref{power}.

\section{Discussion \& Conclusions} \label{Conclusions}

We have presented a series of cosmological \nbody\ simulations which
made use of the hydrodynamic approach to the evolution of structures
(Dom{\'\i}nguez 2000).
This approach is novel in that it deals with the mass density and
velocity fields with explicit account of the coarse-grained nature
intrinsic to any approach of solving, for instance, Poisson's equation via
Monte Carlo sampling of phase-space. This \nbody\ approach unavoidably introduces
finite resolution effects and
there have been systematic studies of the consequences in the context
of cosmological structure formation (Kuhlman, Melott \& Shandarin
1996; Splinter \ea 1998; Moore \ea 1998; Knebe \ea 2000; Power \ea
2003). \nbody\ simulations invariably neglect the dynamical effect of
sub--resolution degrees of freedom altogether. For the first time we
have run simulations including a physical model of the coupling to
the neglected scales.
\nbody\ codes are usually viewed as integrators of the Vlasov--Poisson
system of equations. However, we have argued how grid--based \nbody\ 
codes such as \mlapm\ can be reinterpreted to integrate
hydrodynamic--like equations for the mass density and velocity fields.

The additional, correction term introduced in the hydrodynamic
approach is proportional to a "coupling constant" $B$ which depends
on the smoothing window used to calculate the coarse-grained
fields. It is found to be $B=1/4$ for the triangular--shaped--cloud
window used throughout the \nbody\ code \mlapm. In order to get a
better understanding of the effects of the correction term onto the
evolution of cosmic structures we also performed a simulation with a
higher value $B=1$ --- this later model is not physically
  motivated but rather serves as an "academic toy model" for
  comparison.
The standard
\LCDM\ simulation can be understood as another HAPPI run with the
value $B=0$. In order to allow for a better comparison with other feasible
alternatives to the concordance \LCDM\ model as well as to better
gauge the influence of the correction term, we also simulated the
evolution of the same structures in a \LWDM\ universe.

In this work we concentrated on the comparison of the four simulations
at redshift $z=0$. We analyzed the resulting structures in two
complementary manners: global properties of the mass density and
velocity fields, on the one hand, and specific properties of DM halos,
on the other hand. 
We find appreciable differences between the $B\neq 0$ runs and the
reference ($B=0$) \LCDM\ run, even though the force due to the correction
terms are for most particles one or even two orders of magnitude
smaller than the total force (cf.\ \Fig{HAPPIdens}).
Most remarkably, the correction term favors the proliferation of
low--mass halos, giving the mass distribution a more "grainy"
aspect, as well as the gain of angular momentum specially by low--mass
halos, which also shows up in a velocity field with a larger
vorticity. These effects are quantitatively larger as the value of $B$
increases; for $B=1/4$ the differences lie at the $(10-20)\%$ level
(and even higher for the specific angular momentum at low masses).
A feature in which the $B=1/4$ and $B=1$ runs exhibit an opposite
trend with respect to the $B=0$ run is the concentration of high--mass
halos: the $B=1$ run results in an overabundance of high--mass halos
with a lower concentration 
This is
paralleled by a smaller circular velocity of these halos, and by the
relative lack of power in the spectrum of density fluctuations at
sufficiently small scales, so that the maximum density reached in the
$B=1$ run is much smaller than in the other runs. The $B=1/4$ run,
however, shows precisely the opposite tendency with respect to the
reference run.
One can conjecture that this discrepancy between the $B=1/4$ and $B=1$
runs lies in a difference in the rate of shear and vorticity
generation and of kinetic energy drainage by the correction term. A
comparative study of the structures at different redshifts is required
in order to obtain more precise conclusions about this issue.

The relatively small quantitative differences between the
  $B=1/4$ and the $B=0$ runs evidenced in the properties that we have
  measured suggest that the $B=1/4$ correction term could be
  considered a small perturbation to the $B=0$ evolution. By contrast,
  the results of the $B=1$ run indicate that the correction term
  should not be treated as a perturbation in this case.

Our results agree with the theoretical expectation for the qualitative
behavior of the correction term, which models the effect of
small--scale tidal torques and shear stresses (Dom{\'\i}nguez 2000,
2002; Buchert \& Dom{\'\i}nguez 2005). We observed that the
  correction term is dominant preferentially in walls at high
  redshifts, 
and later on in filaments,
regions of mass accretion onto halos, halo centers
as well as in regions of
  particular dynamical activity (e.g., mergers), that is, regions of
  large gradients in the fields, in concordance with the form of the
  correction term~(\ref{correction}).
The term is expected to act as a drain of
kinetic energy in collapsing regions: this can explain the formation of
small clusters of particles, which would otherwise fly by each other ---
%
instead,
  they can be gravitationally confined by a potential well that is lower than in the $B=0$ model. 
  This could explain the 
low mass halo proliferation for  $B = 1/4 $ or $B = 1$,
as well as the
observed tendency of halos to attain a
  slightly more concentrated configuration in the case of $B=1/4$, when
  the correction term can be considered a small perturbation. If
  $B=1$, on the other hand, the loss of kinetic energy is apparently
  so 
  important
  that, in some cases of halos in regions of high
  dynamical activity, dynamical relaxation and coalescence are slowed
  down noticeably, leading to a multiple-core structure.
  These not completely relaxed halos would then have a lower
  concentration and a lower mass than their LCDM counterparts,
  similarly to the simulation results.


We further confirmed explicitly that the
  correction term acts as a source of vorticity. This relates directly to
  the gain in angular momentum of
  halos, which tends to increase with growing value of $B$, especially at
  the low--mass end of the halo distribution.

Finally, we remark that our findings agree qualitatively with conclusions
following from a comparative study of identical initial
conditions evolved at different resolutions. We have run a series of test simulations where we either switched on the HAPPI correction term or increased the actual force resolution; both methods lead to comparable results that are in qualitative agreement with the conclusions presented here. For a more quantitative analysis we though refer the reader to a future paper in preparation where we will investigate the relationship between HAPPI simulations and higher-resolution ones in more detail. The proliferation
of small halos with increasing resolution has also been reported by other authors in and around
(massive) halos (Klypin et al. 1999, Moore \ea 1998) as well as in voids (Gottl\o ber \ea
2003).  
Concerning the generation of angular momentum though, the
relevance for the formation of realistic disk galaxies has yet to be
determined but there are clear indications that this task requires
good mass and force resolution 
(Governato~\ea 2004). In conclusion, the HAPPI
  implementation seems indeed to be qualitatively consistent with what
  one expects from higher resolution simulations and hence may provide a framework for a better understanding of resolution effects in \nbody\ simulations. However, further work is required to substantiate this possibility.
%

\section*{Acknowledgments}
AK acknowledges funding through the Emmy Noether
Programme by the DFG (KN 755/1).
AD acknowledges funding by the Junta de Andaluc{\'\i}a (Spain) through
the program ``Retorno de Investigadores''.
This work was partially supported  by the MCyT and MEyD (Spain) through
grants AYA-0973 and AYA-07468-C03-03 from the PNAyA.\\
The simulations presented in this paper were carried out on
Swinburne's Centre for Astrophysics~\& Supercomputing Beowulf cluster.


\appendix

\section{Conservation of energy, momentum and angular
  momentum}
\label{sec:conservation} 

The coupling to the small--scales modelled by ${\bf C}$ in
\Eqs{hydrosmallk} injects energy into (or drains energy from) 
the resolved spatial scales. Given the simulation
box $V_{\rm box}$ with periodic boundary conditions, we can define the
total peculiar kinetic energy and the total peculiar mean--field
potential energy as follows:
\begin{eqnarray}
  \label{eq:KU}
  K & = & \frac{1}{2} a^3 \int_{V_{\rm box}} 
  d\bx \; \varrho \, \bu^2 ,\\
  U^{\rm mf} & = & \frac{1}{2 \varrho_b} a^2 \int_{V_{\rm box}} d\bx d{\bf y} \; 
  [\varrho(\bx)-\varrho_b] [\varrho({\bf y})-\varrho_b] \, 
  {\cal S}(\bx - {\bf y}) , \nonumber 
\end{eqnarray}
where the time--independent, symmetric kernel ${\cal S}(\bx)$ is the
solution of the problem
\begin{equation}
  \nabla^2 {\cal S} = 4\pi G (a^3 \varrho_b) \delta_{\rm Dirac}(\bx)  
  \qquad (\bx \in V_{\rm box}).
\end{equation}
The mean--field gravitational acceleration is given by 
\begin{equation}
  \bw^{\rm mf} (\bx) = \mbox{} - \frac{1}{\varrho_b a^2} \nabla_\bx 
  \int_{V_{\rm box}} d{\bf y} \; [\varrho({\bf y})-\varrho_b]
  {\cal S}(\bx-{\bf y}) .
\end{equation}
Then, it is easy to show from \Eqs{hydrosmallk} that the total
peculiar energy ${\cal H} = K + U^{\rm mf}$ satisfies the evolution
equation
\begin{equation}
  \label{eq:newLI}
  \frac{d{\cal H}}{dt} = - H (2 K + U^{\rm mf}) + 
    a^3 \int_{V_{\rm box}} d\bx \; \varrho \bu \cdot {\bf C} .
\end{equation}
This is a generalization of the Layzer--Irvine equation. Due to the
correction term, the condition of ''mean--field virialization'', $2 K
+ U^{\rm mf} = 0$, does not imply a time--independent ${\cal H}$.
The quantity
\begin{equation}
  \label{eq:defI}
  {\cal I} = a {\cal H} + \int da \; K
\end{equation}
is conserved by the original Layzer--Irvine equation but is not
constant according to the generalized equation~(\ref{eq:newLI}).

Concerning momentum and angular momentum, \Eqs{hydrosmallk} do not
violate global conservation. Let $V(t)$ denote a time--dependent
volume defined by the condition that the mass enclosed is constant,
i.e., a Lagrangian domain.
The
peculiar momentum of the domain,
\begin{equation}
  \label{eq:defP}
  {\cal P}_V = a^3 \int_{V(t)} d\bx \; \varrho \bu ,
\end{equation}
verifies the evolution equation
\begin{equation}
  \label{eq:momentum}
  \frac{d{\cal P}_V}{dt} = - H {\cal P}_V + 
    a^3 \int_{V(t)} d\bx \; \varrho (\bw^{\rm mf} + {\bf C}) .
\end{equation}
The correction term can be written as the divergence of a tensor
(Buchert \& Dom{\'\i}nguez 2005)
\begin{displaymath}
  \varrho \, C_i = B L^2 \partial_j \left[
    \varrho (\partial_i w_j^{\rm mf}) + 
    2\pi G a \varrho^2 \delta_{ij} 
    - {1 \over a} \varrho (\partial_k u_i) 
    (\partial_k u_j)  
  \right] ,
\end{displaymath}
so that its contribution in \Eq{eq:momentum} is a surface integral
over the border of $V(t)$. In particular, when $V(t)=V_{\rm box}$,
this surface integral vanishes by periodic boundary conditions and,
since the contribution by $\bw^{\rm mf}$ also vanishes in this case,
\Eq{eq:momentum} states that $a {\cal P}_{V_{\rm box}}$ is a constant
of motion.

In the same manner, one defines the angular momentum of the domain
$V(t)$ with respect to its center of mass ${\bf X}_{\rm cm}(t)$,
\begin{equation}
  \label{eq:defL}
  {\cal L}_V = a^4 \int_{V(t)} d\bx \; (\bx-{\bf X}_{\rm cm}) 
  \times \varrho \bu .
\end{equation}
The evolution equation for this quantity is
\begin{equation}
  \label{eq:angular}
  \frac{d{\cal L}_V}{dt} = a^4 \int_{V(t)} d\bx \; 
  (\bx-{\bf X}_{\rm cm}) 
  \times \varrho (\bw^{\rm mf} + {\bf C}) .
\end{equation}
The contribution by ${\bf C}$ can be written again as a surface
integral over the border of $V(t)$. Thus, when $V(t)=V_{\rm box}$,
\Eq{eq:angular} predicts that ${\cal L}_{V_{\rm box}}$ is also a
constant of motion.

As discussed in Sec.~\ref{HAPPI}, the correction ${\bf C}$ is a source
of vorticity in the otherwise curl--free flow of the ''dust model''.
\Eq{eq:angular} shows that the correction also affects the evolution
of the angular momentum of a domain. 
In this case, however, the effect may not be so noticeable, since
already at the level of the ''dust model'' there are tidal torques by
the mean--field gravity $\bw^{\rm mf}$. 
Moreover, since the contribution by ${\bf C}$ is a surface integral,
it may be expected to be less relevant for a larger domain $V(t)$.
Actually, we can rewrite the definition~(\ref{eq:defL}) by inserting
the identity $2 (\bx-{\bf X}_{\rm cm}) = \nabla |\bx-{\bf X}_{\rm
  cm}|^2$ as
\begin{eqnarray}
  {\cal L}_V & = & \frac{1}{2} a^4 \oint_{\partial V} d{\bf S} \times
  \varrho \bu \, |\bx-{\bf X}_{\rm cm}|^2 - \mbox{} \\
  & & \frac{1}{2} a^4 \int_{V} d\bx \; |\bx-{\bf X}_{\rm cm}|^2 
  \, [\varrho \bomega - (\nabla \varrho) \times \bu ] , \nonumber
\end{eqnarray}
and we see that vorticity is but one contribution to the angular
momentum of a Lagrangian domain.

\section{Minkowski functionals of a Poisson distribution}
\label{sec:poissonMF}

\begin{figure}
  \centerline{\psfig{file=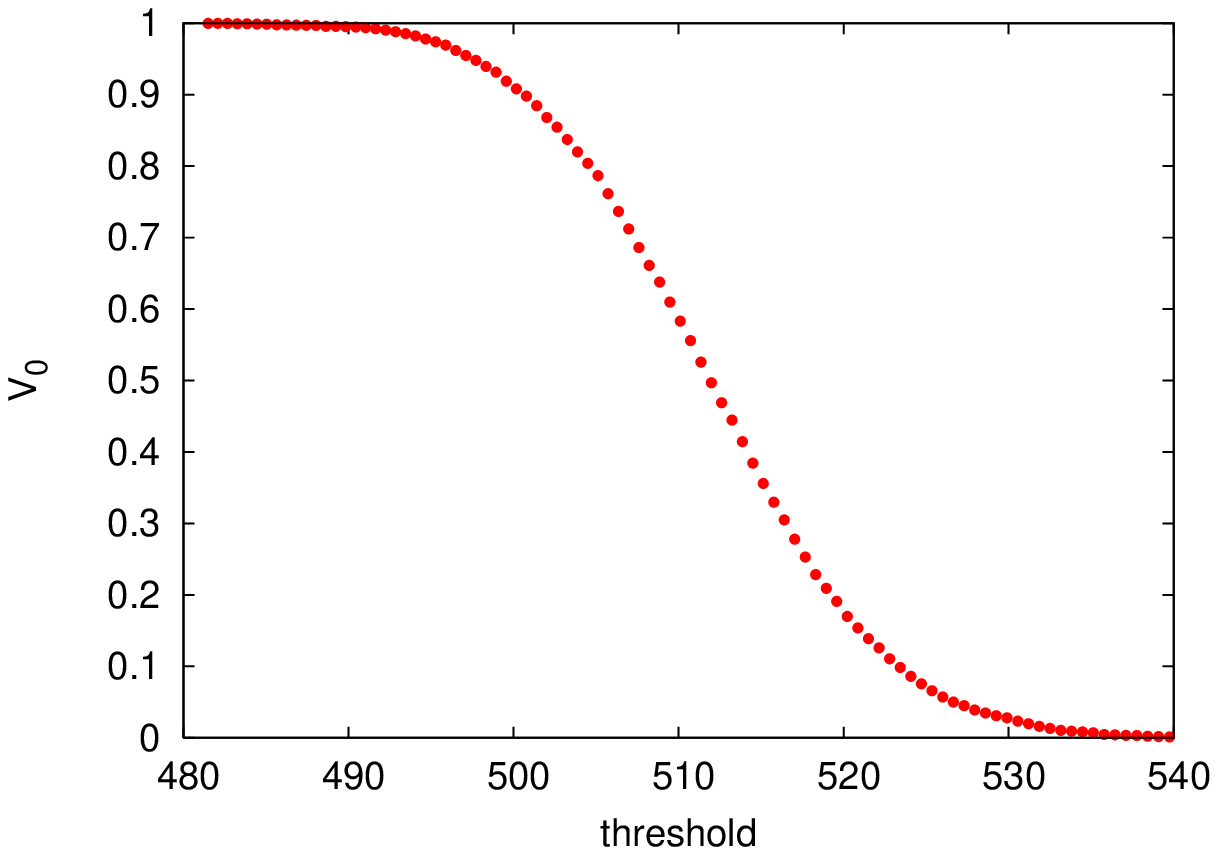,width=.45\hsize}\psfig{file=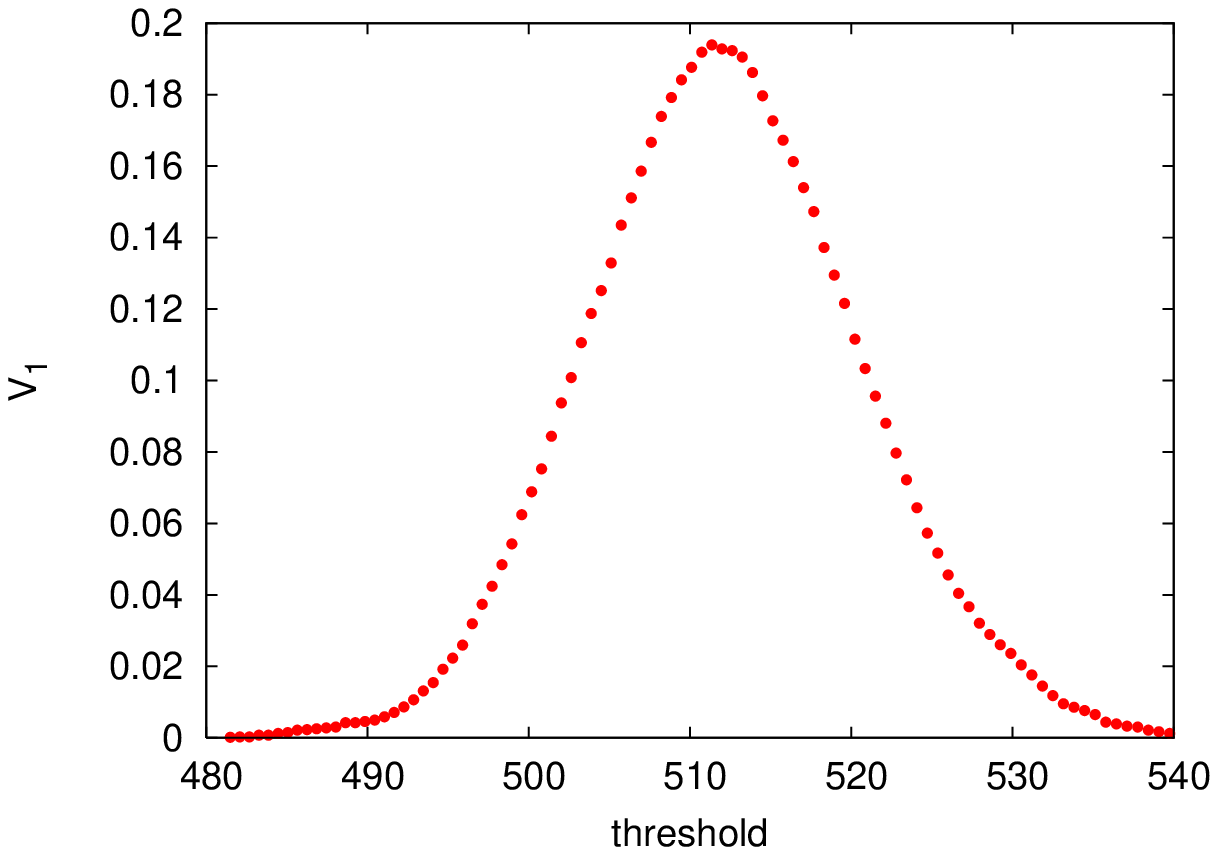,width=.45\hsize}}
  \centerline{\psfig{file=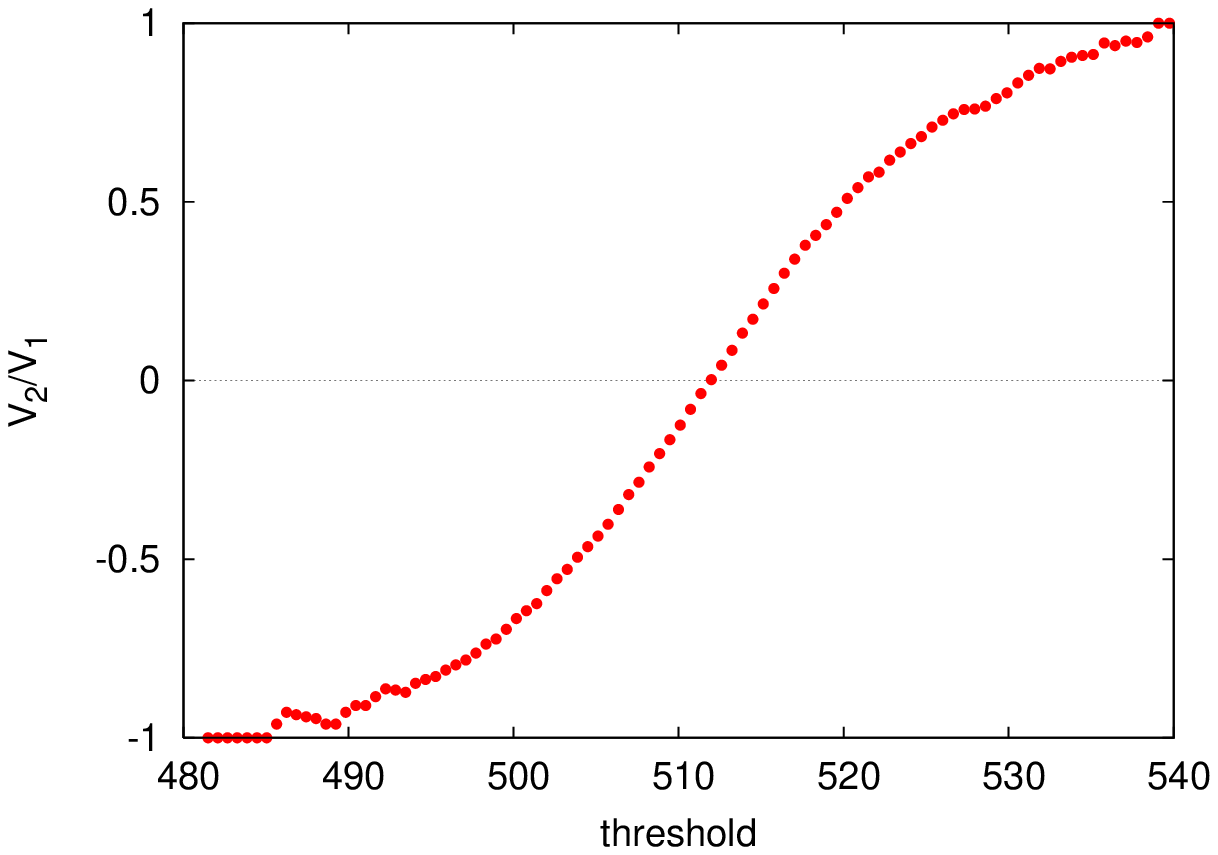,width=.45\hsize}\psfig{file=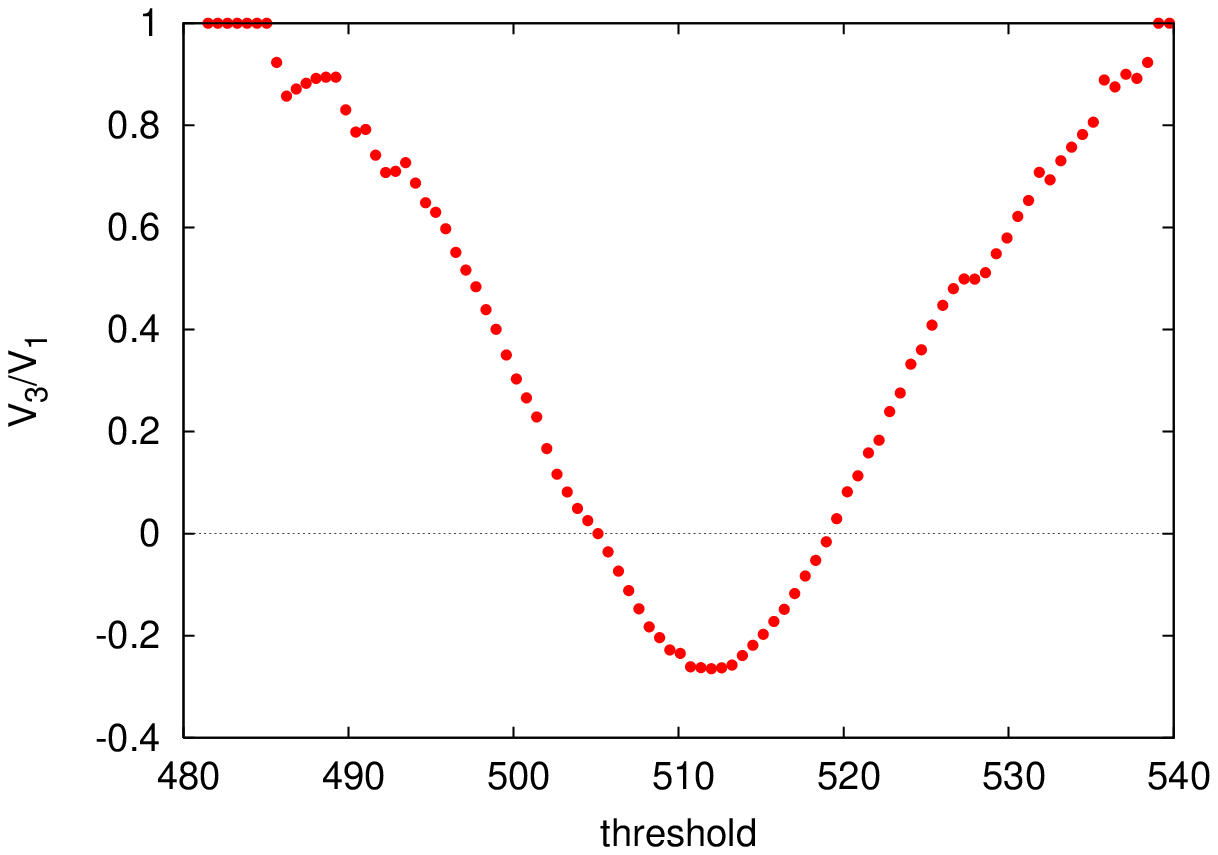,width=.45\hsize}}
  \caption{Minkowski functionals vs.~density threshold of the
    realization of a Poissonian distribution of points.}
  \label{fig:MFpoisson}
\end{figure}

As an illustration of the dependence of the MFs on the threshold,
Fig.~\ref{fig:MFpoisson} shows the MFs of a realization of a Poisson
distribution of points: $128^3$ particles were distributed randomly in
a cubical box, and the density field $\varrho(\bx)$ was obtained by
smoothing with the window~\Eq{TSC} in a cubic grid of $16^3$ nodes.
The plots are symmetric about the mean value of the density ($512$
particles per node) and span a width $\approx$ rms density
($=\sqrt{512}$) along the threshold axis. 
\begin{itemize}
\item The volume $V_0(\thr)$ decreases monotonically as the threshold
  is increased and the high--density regions ($\varrho>\thr$)
  shrink.
\item The area $V_1(\thr)$ first increases as the low--density regions
  ($\varrho<\thr$) expand and, after reaching a maximum, it decreases
  as the high--density regions ($\varrho>\thr$) shrink.
\item The average mean curvature $V_2(\thr)$ increases monotonously
  from a negative value (${\cal S}$ is concave towards the shrinking
  high--density region) to a positive value (${\cal S}$ is convex
  towards the shrinking high--density region).
\item Finally, the genus $V_3(\thr)$ is positive when ${\cal S}$ looks
  bubble--like: there are many unconnected expanding low--density
  regions (``holes'') at small $\thr$, and many unconnected shrinking
  high--density regions (``clusters'') at large $\thr$. $V_3$ is
  negative when ${\cal S}$ is predominantly saddle--shaped: one
  observes many intertwined high-- and low--density regions
  (``tunnels'') at intermediate $\thr$.
\end{itemize}

\end{document}